\documentclass[fontsize=11pt]{article}
\usepackage[natbibapa]{apacite}
\makeatletter
\NAT@longnamesfalse
\makeatother
\AtBeginDocument{%
}

\usepackage{amsmath,amsthm}
\usepackage{xcolor}
\usepackage{enumerate}
\usepackage{graphicx} 
\usepackage{mathtools}
\usepackage{caption,subcaption}
\usepackage{authblk}
\usepackage{multirow}
\usepackage{tablefootnote}
\usepackage{comment}
\usepackage{algorithm,algpseudocode}
\usepackage{url}

\title{\bf\Large Stratified Regression Analysis of Zero-Truncated Recurrent Event Data}

\author[1]{\bf\normalsize Anqi A. Chen \thanks{email: aca142@sfu.ca}}
\author[1]{\bf X. Joan Hu \thanks{email: joanh@stat.sfu.ca}}
\author[2]{\bf Rhonda J. Rosychuk \thanks{email: rhonda.rosychuk@ualberta.ca}}
\affil[1]{\normalsize Department of Statistics and Actuarial Science, Simon Fraser University, Burnaby, British Columbia, Canada}
\affil[2]{Department of Pediatrics, University of Alberta, Edmonton, Alberta, Canada}

\begin{document}



\maketitle

\begin{abstract}
This paper is motivated by an ongoing pediatric mental health care (PMHC) program in which records of mental health-related emergency department (MHED) visits are extracted from population-based administrative databases. A particular interest of this paper is to understand how the visit occurrence depends on the occurrences in the past in a general population. Only information on subjects experiencing MHED visits is available within a subject-specific time window. Thus, the MHED visits may be viewed as zero-truncated recurrent events. Some population census information can be utilized as supplementary information on the covariates of subjects without MHED visits during the study period. We consider an innovative stratified Cox regression model, which is an intensity-based model but requiring only a summary of the event history. We propose an estimation procedure with zero-truncated data integrated with some supplementary information. We establish the consistency and asymptotic normality of the proposed estimator. The finite-sample properties of the estimator are evaluated by simulation, which demonstrates improved performance of the proposed estimator over the maximum likelihood estimator based on zero-truncated data only. We use the PMHC program to illustrate the proposed approach throughout the paper.



\end{abstract}


\noindent{\it keywords:}
Conditional intensity function;
Doubly censored recurrent event data; 
Health administrative data;
Partially known stratification;
Supplementary information.


\section{Introduction}\label{sec:Introduction}


Most mental health issues originate in childhood or adolescence, yet only one out of five children in Canada receive adequate services and treatments \citep{CMHA}. Those with limited access to mental health care turn to emergency departments (EDs) for urgent assistance during crisis \citep{Newton2012cmaj,Cappelli_Cloutier2019,Sima2016}. Our study is motivated by an ongoing pediatric mental health care (PMHC) program, which utilizes population-based administrative data to analyze mental health-related emergency department (MHED) visits made by Alberta residents under 18 years old. Administrative data provide a convenient and cost-effective way to access rich data sources, often at a population level, over an extended observation period 
 \citetext{\citealp{admin_data_2014}; \citealp*{Mason2020}; \citealp{admin_data2015}}. 
However, the use of administrative data poses some challenges since they are not collected for research purposes. For example, they may lack  information on individuals not receiving medical services 
 and protect certain personal details, like birthdates and full postal codes, due to privacy restrictions \citep{admin_data_2014,Hu_Rosychuk2016}.  

In our study, the available data include MHED records and demographic information of Alberta residents under 18 years old experiencing at least one MHED visit between April 1, 2010 and March 31, 2017. We formulate the visits as recurrent events 
and refer to the individuals in the dataset as the MHED cohort. 
When the target population is a population with a minimum of one event, the MHED cohort can be considered as a representative sample of the target population. In this case, the recurrent event data are doubly censored, meaning they are both right- and left-censored. Doubly censored time-to-event data have been analyzed under various models, including a simple linear model by \cite{double_censoring_Cun},  semiparametric transformation models by \cite{double_censoring_Cai} and \cite{double_censoring_LiShuwei2018}, a quantile regression model by \cite{double_censoring_Ji_2012}, a shared-frailty model by \cite{double_censoring_SuYu-Ru2016}, a Cox-Aalen model by \cite{CoxAalenShenPao}, and a proportional hazards model for competing risks data by \cite{competingRisksDoublyCensor_Sankaran}. \cite{Hu_Rosychuk2016} conducted a marginal regression analysis of doubly censored recurrent event data, specifically addressing the case where the censoring times were only known to fall within an interval. Some researchers referred to elapse time between two related events (e.g., virus infection and subsequent disease onset in disease progression) subject to censoring as doubly-censored data \citetext{\citealp*{def2_book_SunJianguo2006TSAo,def_censoring_WangPeijie2018Asrc}; \citealp{def2_censoring_LiShuwei2020}; \citealp*{def2_censoring_WongKinYau2023Srao}}. We consider the first definition. 
When the target population is a general population including subjects without events, the recurrent event data are subject to zero-truncation. That is, the covariates and observation periods are only available for individuals experiencing at least one event; we even do not know the existence of subjects with no events from the dataset. There are relatively fewer studies on zero-truncated recurrent event data. \cite{Hu&Lawless1966_supinf_rateMean} developed nonparametric methods to estimate mean and rate functions for zero-truncated recurrent event data with at least approximate information on the population size and the distribution of observation times across all units including those without events. \cite*{Yi_Hu_Rosychuk2020} adapted the approach of \cite{Hu&Lawless1966_supinf_rateMean} for marginal analysis of zero-truncated recurrent event data under an extended Cox regression model. 
Methods for integrating individual-level data with external aggregate data have also been explored in other studies to enhance efficiency in analyzing small samples of survival data \citetext{\citealp*{huang_qin_Tsai}; \citealp{huang_Qin_2020}; \citealp*{Ping_ding_Wang}; \citealp{AD-ILD_huang2021}}.

Our research goal is to understand how the event occurrence depends on the occurrences in the past. This goal motivates us to consider an intensity-based model rather than a marginal model. Intensity-based models can fully specify a counting process while they require a strong model assumption that the entire event history is known. In contrast, marginal models partially specify a counting process whereas they can provide an easy comparison between groups of subjects and do not depend on the event history \citep{Cook.Lawless.2007}. Rich literature has been established for the analysis of recurrent events. \cite{AGmodel} extended the Cox proportional hazards model \citep{Cox_reg_model}, originally developed for analyzing lifetime data, to the intensity function framework of a counting process for recurrent event data. \cite*{PWP1981} introduced two classes of stratified Cox-type regression models, both with unspecified baseline intensity functions; one class has the baseline as a function of time since the beginning of the study, while the other uses time since the last event. More recent papers considering intensity-based models are \cite{intensity_HuangChiung-Yu2004JMaE}, \cite{intensity_MiloslavskyMaja2004Reai}, \cite{intensity_XuJiajun2017Seot}, \cite{intensity_JazićIna2019Daao}, \cite*{intensity_MaChenchen2021Saoz} and \cite*{intensity-zhao-2022}. The approaches based on 
 marginal models were studied by \cite{rate_PEPEMS1993SGDA}, \cite{rate_LAWLESSJF1995SSRM}, \cite{rate_COOKRICHARDJ.1997MAOR}, \cite{rate_LinD.Y.2000Srft}, \cite{rate_SchaubelDouglasE.2010Eteo}, 
 \cite{marginal_LiShanshan2016Reda},
 \cite{rate_LeeJooyoung2019Dmfm}, \cite{rate_HuangMing-Yueh2023Iseo}, and \cite*{marginal-Xu2024}. Since we lack the full event history for individuals born before the study, we desire something in between conventional intensity-based and marginal models.

This paper makes three contributions. First, we introduce an intensity-based model with stratification that requires only a summary of the event history to fully specify the counting process. Secondly, we propose an estimation procedure of integrating zero-truncated recurrent event data with some supplementary information under the innovative model. 
Last but not least, this paper can bring some practical insights on how the past MHED visits influenced the occurrence of the later visits. The rest of the paper is organized as follows. We introduce the notation and modeling in Section~\ref{sec:notation_model}.
Sections~\ref{sec:estimation_truncated} and \ref{sec:estimation_censoring} show the estimation procedures for analyzing zero-truncated recurrent event data, both independently and integrated with some supplementary information. Each section starts with a relatively straightforward procedure with a fully known stratification variable in the models and then we adapt the procedures to accommodate a partially known stratification variable. 
Additionally, we establish the asymptotic properties of the resulting estimators. In Section~\ref{sec:simulation}, we conduct a simulation study  to examine the numerical performance of the approaches introduced in Sections~\ref{sec:estimation_truncated} and \ref{sec:estimation_censoring}. In Section~\ref{sec:real_data}, we apply the approaches to the MHED data. Section~\ref{sec:conclusion} provides some final remarks and discussions.

\section{Notation and Modeling}\label{sec:notation_model}
We are interested in a general population under age $A^\star$, say 18 year-old, denoted by $\mathcal{P}=\mathcal{P}_1\cup \mathcal{P}_0$, where $\mathcal{P}_1$ represents one sub-population experiencing at least one event and $\mathcal{P}_0$ is the other sub-population without events before age $A^\star$.  We assume that subjects in $\mathcal{P}$ are independent with each other. Let $N_i(a)$ denote the number of events made by subject $i$ up to age $a$ since birth, for $i\in \mathcal{P}$ and $0\leq a <A^\star$. By convention, $N_i(0)\equiv 0$. The history information of subject $i$ is defined as $\mathcal{H}_i(a)=\big\{N_i(u): 0\leq u<a\big\}$. We consider a stratification variable
$S_i(a)=S\big\{\mathcal{H}_i(a)\big\}$ with a finite number of values for a fixed $a$; for $a \in [0, A^\star)$, $S_i(a)$ is a left-continuous and non-decreasing function. Let $Z_i$ represent time-independent covariates of subject $i$. Consider the conditional intensity function of
the counting process $N_i(\cdot)=\big\{N_i(a):a\geq 0\big\}$ to be, with $S_i(a)=s$,
\begin{equation}
\lambda(a\mid \mathcal{H}_i(a), Z_i)
=\lambda_{0s}(a) \exp\{\beta_s^{'} Z_i\}\text{, }a>0.
\label{eq:model}
\end{equation} 
An example of the stratification variable is 
\begin{align}
    S_i(a)=\begin{cases}1, & N_i(a-) = 0\\2, &N_i(a-) > 0\end{cases}.
\label{eq:stratification_variable}
\end{align}
This indicates that subject $i$ is in stratum 1 if they have not experienced their first event before age $a$; otherwise, they are in stratum 2.

Our statistical goal is to estimate the unspecified baseline intensity functions $\{\lambda_{0s}(\cdot):s\in\mathcal{S}\}$  and the regression coefficients
 $\{\beta_s:s\in\mathcal{S}\}$, where $\mathcal{S}$ is the set of
all the possible values of $S_i(a)$. Table \ref{Tab:submodels} shows relevant model specifications for Model (\ref{eq:model}). In the model labels, the first and second letters represent whether stratification (S) or no stratification (N) is applied to the baseline and the coefficients, respectively. The final letter indicates whether the baseline is time-independent (C) or time-varying (V). Model (NNC) represents a Poisson process and Model (SSV) corresponds to the proposed model.

\begin{table}[ht!]
\centering
\caption{Relevant model specifications}
\label{Tab:submodels}
\small
\vspace{-0.25cm}
\begin{tabular}{rcc}
\hline
\hline
Model & Baseline & Coefficient\\
  \hline
(NNC) & $\lambda_{0s}(a)=\lambda_0$ & $\beta_s=\beta$ \\ 
(NSC) & $\lambda_{0s}(a)=\lambda_0$ & $\beta_s=\beta_s$\\
(SNC) & $\lambda_{0s}(a)=\lambda_{0s}$ & $\beta_s=\beta$\\
(SSC) & $\lambda_{0s}(a)=\lambda_{0s}$ & $\beta_s=\beta_s$\\ 
 (NNV) & $\lambda_{0s}(a)=\lambda_0(a)$ & $\beta_s=\beta$\\ 
 (NSV) & $\lambda_{0s}(a)=\lambda_0(a)$ & $\beta_s=\beta_s$\\ 
 (SNV) &  $\lambda_{0s}(a)=\lambda_{0s}(a)$ & $\beta_s=\beta$\\ 
 (SSV) &  $\lambda_{0s}(a)=\lambda_{0s}(a)$ & $\beta_s=\beta_s$\\ 
   \hline
   \hline
\end{tabular}
\end{table}


Let $\mathcal{O}$ represent a random sample of $\mathcal{P}$ within a predetermined data extraction window, $[W_L, W_R]$, in the calendar time. The sample $\mathcal{O}$ consists of two subsets $\mathcal{O}_1$ and $\mathcal{O}_0$: $\mathcal{O}=\mathcal{O}_1\cup \mathcal{O}_0$, where subjects in $\mathcal{O}_1$ have at least one event and subjects in $\mathcal{O}_0$ have no events within $[W_L, W_R]$. Let $B_i$ denote the birthdate of subject $i$ in the calendar time. The event times for subject $i \in \mathcal{O}_1$
are observed over $(C_{L_i}, C_{R_i}]$ in the personal time (that is, age in years), where
$C_{L_i}=\max(0, W_L-B_i)$ and
$C_{R_i}=\min(A^\star,W_R-B_i)$. We consider the left open interval because the information up to $C_{L_i}$ is unavailable to us. We assume that both $W_L$ and $W_R$ are independent of the events,
and subject $i$'s birthdate $B_i$ is independent of
the counting process $N_i(\cdot)$.

The total number of events made by subject $i$ over the subject-specific time interval $\big(C_{L_i},C_{R_i}\big]$ 
 is denoted by $N^{\star}_i =N_i(C_{R_i})-N_i(C_{L_i})$; $N^{\star}_i>0$ for all  $i \in \mathcal{O}_1$. Let $T_{ij}$ be the calendar time of the $j$th
observed event made by subject $i$ and then the age of subject $i$ at the $j$th event time
is $A_{ij}=T_{ij}-B_i$
for $i\in \mathcal{O}_1$ and $j=1,\ldots, N^{\star}_i$. The available data are represented as $\mathcal{Q}_1=\bigcup_{i\in\mathcal{O}_1} \mathcal{Q}_{1i}=\bigcup_{i\in\mathcal{O}_1}\big\{\{dN_i(a):C_{L_i}<a\leq C_{R_i}\}\bigcup_{}^{}\{Z_i\}\big\}$, where $dN_i(\cdot)$ represents the change in $N_i(\cdot)$. With no tie, $dN_i(\cdot)=1$ at the event times $a_{ij}$'s and $dN_i(\cdot)=0$ otherwise. We assume that $\mathcal{O}_1$ is a random sample of $\mathcal{P}_1$, then $\mathcal{O}_0$ is a random sample of $\mathcal{P}_0$. We aim to use available information on $\mathcal{O}_1$ possibly with some additional information to make inference on the general population $\mathcal{P}$. 

\section{Estimation with Zero-Truncated Recurrent \\Event Data} \label{sec:estimation_truncated}
\noindent3.1\textit{\hspace{0.4cm}Estimation When the Stratification Variable is Fully Available}\label{sec:truncation_known_S} 

\noindent Since we lack information on subject $i\in\mathcal{O}_0$, we do not have a representative sample of the general population $\mathcal{P}$ and cannot estimate the baseline intensity functions $\boldsymbol{\lambda}_{0}(\cdot)=\{\lambda_{0s}(\cdot):s\in\mathcal{S}\}$ and the regression coefficients $\boldsymbol{\beta}=\{\beta_s:s\in\mathcal{S}\}$ directly under Model (\ref{eq:model}). One straightforward approach to address this issue is to derive the conditional intensity function for subjects in the sub-population $\mathcal{P}_1$
 with a given stratification variable. The induced model is written as
\begin{align*}
\begin{split}
    \lambda^{\star}(a\mid \mathcal{H}_i(a), Z_i)
 =\lambda(a\mid \mathcal{H}_i(a), Z_i)\frac{P\bigl(N^{\star}_i\geq1|dN_i(a)=1, \mathcal{H}_i(a), Z_i\bigr)}{P\bigl(N^{\star}_i\geq1|\mathcal{H}_i(a), Z_i\bigr)}.
\end{split}
\end{align*}
An induced model with the stratification variable (\ref{eq:stratification_variable}) is provided in Appendix \ref{sec:Appendix_Estimation_details}1 as an example. We use the induced model to link the parameters from Model (\ref{eq:model}) so we can directly estimate the $\boldsymbol{\lambda}_{0}(\cdot)$ and $\boldsymbol{\beta}$ under the induced model. For simplicity, we consider a special case of the induced model; that is, let $\lambda_{0s}(a)=\lambda_{0s}$ for $s\in \mathcal{S}$, which corresponds to Model (SSC) in Section \ref{sec:notation_model}.

 Let $Y_i^{(c)}(a)= I\big(a \in (C_{Li}, C_{Ri}]\big)$ and $Y_i^{(s)}(a)=I\big(a: S_i(a)=s\big)$.
 When the stratification variable is fully known, the log-likelihood function 
 is 
\begin{align*}
\begin{split}
\ell(\boldsymbol{\beta}|\boldsymbol{S}(\cdot),\mathcal{Q}_{1})
   &=\sum_{i\in \mathcal{O}_1}^{} \Biggl\{ \sum_{s\in\mathcal{S}}^{}\biggl[\int_{0}^{A^{\star}} Y_i^{(s)}(a)Y_i^{(c)}(a)\log\big(\lambda^{\star}(a\mid \mathcal{H}_i(a), Z_i)\big) dN_i(a)  \\
     &\quad -\int_{0}^{A^{\star}}Y_i^{(s)}(a)Y_i^{(c)}(a)\lambda^{\star}(a\mid \mathcal{H}_i(a), Z_i) da\biggr] \Biggr\}.
\end{split}
\end{align*}
\normalsize
Let $\alpha_s=(\log\lambda_{0s},\beta_s^{'})^{'}$, $Z^\star_i=(1, Z_i^{'})^{'}$, and $\boldsymbol{\alpha}=\{\alpha_{s}:s\in\mathcal{S}\}$. The log-likelihood function can be re-written as $\ell^\star(\boldsymbol{\alpha}|\boldsymbol{S}(\cdot),\mathcal{Q}_{1})$ by replacing  $\lambda_{0s}e^{\beta_s^{'}Z_i}$ in $\ell(\boldsymbol{\beta}|\boldsymbol{S}(\cdot),\mathcal{Q}_{1})$ with $e^{\alpha_s^{'}Z_i^\star}$. Appendix \ref{sec:Appendix_Estimation_details}2 shows an example of $\ell^\star(\boldsymbol{\alpha}|\boldsymbol{S}(\cdot),\mathcal{Q}_{1})$ with the stratification variable (\ref{eq:stratification_variable}). We can use the maximum likelihood estimation to estimate $\boldsymbol{\alpha}$. 
\bigskip

\noindent3.2\textit{\hspace{0.4cm}Estimation When the Stratification Variable is Partially Available} \label{sec:truncation_P_avail_S}

\noindent When the stratification variable is partially known to us, we use the expectation-maximization (EM) algorithm  \citep*{EM_algorithm} to estimate $\boldsymbol{\alpha}$ under Model (SSC). The EM algorithm is outlined in Algorithm \ref{alg:EM}.

\begin{algorithm}[ht!]
\caption{Estimate $\boldsymbol{\alpha}$}\label{alg:EM}
\begin{algorithmic}
\State Let $\boldsymbol{\alpha}^{(k)}$ denote the estimate of $\boldsymbol{\alpha}$ in the $k$-th iteration for $k=0,1,\cdots$. The $\boldsymbol{\alpha}^{(0)}$ represents the initial values.
\begin{enumerate}
    \item \textit{Expectation step (E-step)}: Given the observed data $\mathcal{Q}_1$ and the current estimate $\boldsymbol{\alpha}^{(k)}$, compute the expected value of $\ell^\star(\boldsymbol{\alpha}|\boldsymbol{S}(\cdot),\mathcal{Q}_{1})$: $E\big[\ell^\star(\boldsymbol{\alpha}|\boldsymbol{S}(\cdot),\mathcal{Q}_{1})\big|$ $\mathcal{Q}_1,\boldsymbol{\alpha}^{(k)}\big]$.

    \item \textit{Maximization step (M-step)}: Find the estimate 
\begin{equation*}
   \widehat{\boldsymbol{\alpha}}^{(k+1)}={\underset {\boldsymbol {\alpha\in \Theta}}{\operatorname {arg\,max} }}~E\big[\ell^\star(\boldsymbol{\alpha}|\boldsymbol{S}(\cdot),\mathcal{Q}_{1})\big|\mathcal{Q}_1,\boldsymbol{\alpha}^{(k)}\big], 
\end{equation*}
where $\Theta$ denotes the parameter space of $\boldsymbol{\alpha}$.
\end{enumerate}
Repeat Steps 1 and 2 until $||\alpha_s^{(k)}-\alpha_s^{(k-1)}||_1/||\alpha_s^{(k-1)}||_1\leq \tau$ for $\forall s \in \mathcal{S}$, where $\tau$ is a predetermined tolerance level.
\end{algorithmic}
\end{algorithm}

The limit of the sequence is denoted by $\widehat{\boldsymbol{\alpha}}^{z}$, which is the maximum likelihood estimator (MLE) of $\boldsymbol{\alpha}$ based on zero-truncated data. The MLE of $\lambda_{0s}$ is the exponential of the first component of $\widehat{\alpha}_s^z$ and the MLE of $\beta_s$ consists of the remaining components in $\widehat{\alpha}_s^z$.

\bigskip

\noindent \textbf{Proposition 1}. \textit{Under some regularity conditions \citep{casella2002statistical}, the MLE $\widehat{\boldsymbol{\alpha}}^z$ has the following asymptotic properties: }

 \begin{enumerate}[(i)]
    \item strong consistency: $\widehat{\boldsymbol{\alpha}}^z \xrightarrow{a.s.} \boldsymbol{\alpha}_0$ as $n_1\rightarrow \infty$, with $n_1=|\mathcal{O}_1|$, and
     \item asymptotic normality: $\sqrt{n_1}(\widehat{\boldsymbol{\alpha}}^z-\boldsymbol{\alpha}_0)\xrightarrow{d} N \big(\mathbf{0},\Pi^z(\boldsymbol{\alpha}_0)^{-1}\big)$ as $n_1\rightarrow \infty$, where $\boldsymbol{\alpha}_0$ represents the true parameters,
\begin{align*}
    \Pi^z(\boldsymbol{\alpha})&=-\frac{1}{n_1}E\Bigg[\frac{\partial^2 log \big(L(\boldsymbol{\alpha}|\mathcal{Q}_1 )\big)}{\partial \boldsymbol{\alpha}^2}\Bigg],
\end{align*}
and
\begin{align*}
    L(\boldsymbol{\alpha}|\mathcal{Q}_1 )=\prod_{i\in\mathcal{O}_1} \bigg\{\sum_{s\in\mathcal{S}}L_s\big(\boldsymbol{\alpha}\big|S_i(\cdot),\mathcal{Q}_{1i}\big)P\big(Y_i^{(s)}(C_{L_i})=1\big|Z_i\big)\bigg\}.
\end{align*}
 \end{enumerate}
The cardinality $|\mathcal{O}_1|$ denotes the number of subjects in $\mathcal{O}_1$. The $L_s\big(\boldsymbol{\alpha}_0\big|S_i(\cdot),\mathcal{Q}_{1i}\big)$ represents the probability of observing $\mathcal{Q}_{1i}$ given subject $i$ is in stratum $s$ at age $C_{L_i}$. Appendix \ref{sec:Appendix_Estimation_details}3 provides additional information on $ L(\boldsymbol{\alpha}|\mathcal{Q}_1 )$.

\section{Estimation with Zero-Truncated Recurrent \\Event Data Augmented with Population Census Information}\label{sec:estimation_censoring}

4.1\textit{\hspace{0.4cm}Estimation When the Stratification Variable is Fully Available} \label{sec:censoring_known_S} 

\noindent The zero-truncated data are population-based and contain information on all subjects experiencing at least one event within the data extraction window, $[W_L, W_R]$. Eligible study subjects excluded from the dataset all have $N_i^\star=0$ and their covariates and observation periods are missing. If we can obtain supplementary information for subjects in $\mathcal{O}_0$, we can address the missing problem and construct a representative sample of the general population $\mathcal{P}$. For example, population census data can provide aggregate geographic and demographic information on Alberta residents across different age groups, including those without MHED visits during the study period. Once integrated with population census data, the recurrent event data are doubly censored, where $C_{Li}$ and $C_{Ri}$ represent the left- and right-censoring times for subject $i$, respectively. In the remainder of this subsection, we start with the estimation procedure for an ideal case where all necessary information is available. We then present the estimation function for the case where covariate information is missing for subjects in $\mathcal{O}_0$ and population census data are available to us.

With a fully known stratification variable, $S_i(a)$, and a random sample of $\mathcal{P}$, represented by $\mathcal{O}$, the log-partial likelihood function of $\boldsymbol{\beta}$ under Model (\ref{eq:model}) is 
\begin{align*}
    &\text{log}\big(PL(\boldsymbol{\beta})\big)\\
    &= \quad \sum_{i\in \mathcal{O}_1}\sum_{s\in\mathcal{S}}^{}\int_{0}^{A^{\star}} Y_i^{(s)}(a)\Bigl\{\beta_s^{'}Z_i -\text{log}\bigl(\sum_{k\in \mathcal{O}} Y_k^{(s)}(a) Y_k^c(a)e^{\beta_s^{'}Z_k}\bigr)\Bigr\}Y_i^{(c)}(a)dN_i(a).
\end{align*}
Let $\bar{Z}_s(\beta_s;a)=G_s^{(1)}(\beta_s;a)/G_s^{(0)}(\beta_s;a)$, where $G_s^{(q)}(\beta_s;a)=\sum_{i\in \mathcal{O}} Y_i^{(s)}(a)\times Y_i^{(c)}(a)$ $Z_i^{\otimes q}\exp\{\beta_s^{'}Z_i\}$ for $q=0, 1, 2$ with $b^{\otimes 0}=1$, $b^{\otimes 1}=b$, and $b^{\otimes 2}=bb^{'}$. Then, the corresponding partial score function of $\beta_s$ is
\begin{align}
    U_{s}(\boldsymbol{\beta})=\frac{\partial \text{log}(PL(\boldsymbol{\beta}))}{\partial \beta_s}
    =\sum_{i\in \mathcal{O}_1}^{}\int_{0}^{A^{\star}}Y_i^{(s)}(a)\big\{Z_i-\bar{Z}_s(\beta_s;a)\big\}  Y_i^{(c)}(a)dN_i(a),
    \label{eq:P_score_fun-ideal}
\end{align}
for $s\in\mathcal{S}$. The solution of $U_{s}(\boldsymbol{\beta})=\mathbf{0}$ is the maximum partial likelihood estimator (MPLE) of $\beta_s$. In addition, Equation (\ref{eq:P_score_fun-ideal}) is an unbiased estimating function of $\beta_s$ under Model (\ref{eq:model}). 

When the covariate information for subjects in $\mathcal{O}_0$ is available only in an aggregate form, we can approximate $U_{s}(\boldsymbol{\beta})$ by 
\begin{align}
\tilde{U}_{s}(\boldsymbol{\beta})=\sum_{i\in \mathcal{O}_1}^{}\int_{0}^{A^{\star}}Y_i^{(s)}(a)\big\{Z_i-\tilde{\bar{Z}}_s(\beta_s;a)\big\}  Y_i^{(c)}(a)dN_i(a),
    \label{eq:P_score_fun-appx}
\end{align}
where $\tilde{\bar{Z}}_s(\beta_s;a)=\tilde{G}_s^{(1)}(\beta_s;a)/\tilde{G}_s^{(0)}(\beta_s;a)$ approximates $\bar{Z}_s(\beta_s;a)$ with 
\begin{equation*}
    \tilde{G}_s^{(q)}(\beta_s;a)=\sum_{z\in \mathcal{Z}}^{}\Bigl\{
P\bigl(Y^{(s)}(a)=1|Z=z\bigr)z^{\otimes q}\exp\{\beta_s^{'}z\}\bigl[\sum_{l}\mathcal{C}(l,z,\lfloor a \rfloor)\bigr]\Bigr\}.
\end{equation*}
The 
$\mathcal{Z}$ represents a collection of all the covariate combinations, and $\mathcal{C}(l,z,\lfloor a \rfloor)$ is the number of subjects at age $\lfloor a \rfloor$ with covariates $Z=z$ in calendar year $l$. Here, we adapt \cite*{Yi_Hu_Rosychuk2020}'s method of integrating population census information with zero-truncated recurrent event data. The expression of $\tilde{G}_s^{(q)}(\beta_s;a)$ implies that all the covariates are discrete with a finite number of values and covariate information on eligible subjects is summarized by calendar year and by integer age in the population census data. If the census data are recorded monthly or daily, we could obtain a better approximation of $U_{s}(\boldsymbol{\beta})$. 

Given the value of $\boldsymbol{\beta}$ and the population census information, the estimating equation of cumulative baseline function $\boldsymbol{\Lambda}_{0}(\cdot)=\{\Lambda_{0s}(\cdot):s\in\mathcal{S}\}$ can be approximated by

\begin{align}
    \tilde{V}_{s}(a\big|\boldsymbol{\beta})=\sum_{i\in \mathcal{O}_1}^{}Y_i^{(s)}(a)Y_i^{(c)}(a)dN_i(a)-\tilde{G}_s^{(0)}(\beta_s;a) d\Lambda_{0s}(a)=0,
    \label{eq:est_baseline_fun-appx}
\end{align}
for $a\in (0, A^\star)$ and $s\in\mathcal{S}$.
We jointly solve $\tilde{U}_{s}(\boldsymbol{\beta})=\mathbf{0}$ for $\beta_{s}$ and Equation (\ref{eq:est_baseline_fun-appx}) for $d\Lambda_{0s}(\cdot)$ for $s\in\mathcal{S}$ since the probability $P\bigl(Y^{(s)}(a)=1|Z=z\bigr)$ potentially includes all the coefficients and baselines. Let $\tilde{\beta}_s$ represent the resulting estimator of $\beta_{s}$.  Integrating the estimator of $d\Lambda_{0s}(\cdot)$ from $0$ to $a$ yields an estimator of the cumulative baseline intensity function under Model (\ref{eq:model}); that is, 
\begin{align}
    \tilde{\Lambda}_{0s}(a\big|\boldsymbol{\beta})=\sum_{i\in \mathcal{O}_1} \int_{0}^{a}\frac{Y_i^{(s)}(u)}{\tilde{G}_s^{(0)}(\beta_s;u)}Y_i^{(c)}(u)dN_i(u).
\end{align}
We let $0/0=0$ by convention. 
Plugging in $\widetilde{\boldsymbol{\beta}}=\{\tilde{\beta}_{s}:s\in\mathcal{S}\}$, we obtain the Breslow estimator \citep{breslow_est1972} $\tilde{\Lambda}_{0s}(a\big|\widetilde{\boldsymbol{\beta}})$ for $s\in\mathcal{S}$. Once we know the estimators of the cumulative baseline functions, we can generate the estimators of the baseline intensity functions, $\tilde{\lambda}_{0s}(a\big|\widetilde{\boldsymbol{\beta}})$'s, by taking the first derivative with respect to $a$. 

\bigskip

\noindent4.2\textit{\hspace{0.4cm}Estimation When the Stratification Variable is Partially Available} \label{sec:censoring_P_avail_S}

\noindent When the stratification variable $S_i(a)$ is only partially known, we do not always know the value of $Y_i^{(s)}(a)$ for $a\in (C_{L_i},C_{R_i}]$. As a result, the approximate partial score function $\tilde{U}_{s}(\boldsymbol{\beta})$ shown in (\ref{eq:P_score_fun-appx}) is not always evaluable. Instead, we replace $Y_i^{(s)}(a)$ with $P\bigl(Y_i^{(s)}(a)=1|\mathcal{Q}_{1i}\bigr)$ and obtain the following two estimating equations for $\beta_s$ and $\Lambda_{0s}(\cdot)$:
\small
\begin{align}
\begin{split}
    \tilde{U}_{s}^{\star}(\boldsymbol{\beta})&=\sum_{i\in \mathcal{O}_1}^{}\int_{0}^{A^{\star}}P\bigl(Y_i^{(s)}(a)=1|\mathcal{Q}_{1i}\bigr)\big\{Z_i-\tilde{\bar{Z}}_s(\beta_s;a)\big\}  Y_i^{(c)}(a)dN_i(a)=\mathbf{0},
    \label{eq:score_fun-p_s_appx}
    \end{split}
\end{align}
\normalsize
and
\small
\begin{align}
\begin{split}
    &\tilde{V}_{s}^{\star}(a\big|\boldsymbol{\beta}) =\sum_{i\in \mathcal{O}_1}^{}P\bigl(Y_i^{(s)}(a)=1|\mathcal{Q}_{1i}\bigr)Y_i^{(c)}(a)dN_i(a)-\tilde{G}_s^{(0)}(\beta_s;a) d\Lambda_{0s}(a)=0.
    \label{eq:est_baseline_fun_P_S-appx}
    \end{split}
\end{align}
\normalsize
We jointly solve the two equations for $\beta_s$ and $\Lambda_{0s}(\cdot)$ for all $s\in\mathcal{S}$, and obtain their estimators, denoted by $\tilde{\tilde{\beta}}_s$ and $\tilde{\tilde{\Lambda}}_{0s}(\cdot)$. The consistency and asymptotic normality of 
$\widetilde{\widetilde{\boldsymbol{\beta}}}=\{\tilde{\tilde{\beta}}_{s}:s\in\mathcal{S}\}$,
which is our main interest, are stated as follows.

\noindent \textbf{Proposition 2}. \textit{Under conditions (I)-(V), the estimator $\widetilde{\widetilde{\boldsymbol{\beta}}}$ has the following asymptotic properties:} 

 \begin{enumerate}[(i)]
    \item strong consistency: $\widetilde{\widetilde{\boldsymbol{\beta}}} \xrightarrow{a.s.} \boldsymbol{\beta}_0$ as $n\rightarrow \infty$, 
    and
     \item asymptotic normality: $\sqrt{n}(\widetilde{\widetilde{\boldsymbol{\beta}}}-\boldsymbol{\beta}_0)\xrightarrow{d} N \big(\mathbf{0},AV(\boldsymbol{\beta}_0)\big)$ as $n\rightarrow \infty$, where $\boldsymbol{\beta}_0$ represents the true parameters and $n=|\mathcal{O}|$.
 \end{enumerate}
The proof for Proposition 2, 
along with the sufficient conditions, is outlined in Appendix \ref{sec:Appendix_proofs}.

To estimate the $AV(\boldsymbol{\beta}_0)$, we apply the following resampling method using Poisson multipliers with mean $1$ \citep*{Poisson_multi} :
\begin{enumerate}
    \item 	Generate a set of multipliers $W^{(b)}=(W_1^{(b)},\cdots,W_{|\mathcal{O}_1 |}^{(b)} )'$, where $W_i^{(b)}\overset{\mathrm{iid}}{\sim}\text{Poisson}(1)$ for $i=1,\cdots, |\mathcal{O}_1 |$. These multipliers are independent of the data.
    \item 	Given a realization of $W^{(b)}$, jointly solve the following equations for $\beta_s$ and $\Lambda_{0s}(\cdot)$:
\begin{align*}
\begin{split}
    \sum_{i\in \mathcal{O}_1}^{}\int_{0}^{A^{\star}}P\bigl(Y_i^{(s)}(a)=1|\mathcal{Q}_{1i}\bigr)\big\{Z_i-\tilde{\bar{Z}}_s(\beta_s;a)\big\}  Y_i^{(c)}(a)dN_i(a) W_i^{(b)}=\mathbf{0},
    \end{split}
\end{align*}
and
\begin{align*}
\begin{split}
        \Lambda_{0s}^{(b)}(a\big|\boldsymbol{\beta})=\int_{0}^{a}\sum_{i\in \mathcal{O}_1} \frac{P\bigl(Y_i^{(s)}(u)=1|\mathcal{Q}_{1i}\bigr)}{\tilde{G}_s^{(0)}(\beta_s;u)}Y_i^{(c)}(u)dN_i(u)W_i^{(b)}.
    \end{split}
\end{align*}
\item Repeat Steps 1 and 2 for $B$ times to obtain $\widetilde{\widetilde{\boldsymbol{\beta}}}^{(1)},\cdots, \widetilde{\widetilde{\boldsymbol{\beta}}}^{(B)}$. The estimated asymptotic variance of $\widetilde{\widetilde{\boldsymbol{\beta}}}$ is the sample variance of the estimates $\widetilde{\widetilde{\boldsymbol{\beta}}}^{(1)},\cdots, \widetilde{\widetilde{\boldsymbol{\beta}}}^{(B)}$. 
\end{enumerate}
The resampling approaches proposed by \cite*{Poisson_multi} utilize centered Poisson multipliers with variance 1, whereas our approach does not center the Poisson multipliers.  This difference is addressed in the two equations in Step 2.  Additionally, the multipliers in Step 1 can alternatively be drawn from a standard normal distribution, as suggested by \cite*{Lin1993_normal_multi}.

Since both $P\bigl(Y_i^{(s)}(a)=1|\mathcal{Q}_{1i}\bigr)$ and $P\bigl(Y^{(s)}(a)=1|Z=z\bigr)$ may involve all the baseline functions and the regression coefficients $\bigl($see Appendix \ref{sec:Appendix_Estimation_details}4 for an example of $P\bigl(Y_i^{(s)}(a)=1|\mathcal{Q}_{1i}\bigr)\bigr)$, in practice, it is computationally challenging to jointly solve the estimating equations (\ref{eq:score_fun-p_s_appx}) and (\ref{eq:est_baseline_fun_P_S-appx}). Therefore, we propose Algorithm \ref{alg:approch_2} to estimate $\boldsymbol{\lambda}_{0}(\cdot)$ and $\boldsymbol{\beta}$ in a more practical manner.

\bigskip
\begin{algorithm}[ht!]
\caption{Estimate $\boldsymbol{\lambda}_{0}(\cdot)$ and $\boldsymbol{\beta}$}\label{alg:approch_2}
\begin{algorithmic}
\State Let $\boldsymbol{\lambda}_{0}^{(r)}(\cdot)$ and $\boldsymbol{\beta}^{(r)}$ for $r=0,1,\cdots$ denote the estimates of the baseline intensity functions and the coefficients in the $r$-th iteration. The $\boldsymbol{\lambda}_{0}^{(0)}(\cdot)$ and $\boldsymbol{\beta}^{(0)}$ represent predetermined initial functions and values.
\begin{enumerate}
    \item Given $\boldsymbol{\lambda}_{0}^{(r)}(\cdot)$ and $\boldsymbol{\beta}^{(r)}$, estimate $P\bigl(Y_i^{(s)}(a)=1|\mathcal{Q}_{1i}\bigr)$ for $i\in \mathcal{O}_1$ and $P\bigl(Y^{(s)}(a)=1|Z=z\bigr)$ for $z\in\mathcal{Z}$. Then, solve $\tilde{U}_{s}^{\star}(\boldsymbol{\beta})=\mathbf{0}$ for $\beta_s$ to obtain $\boldsymbol{\beta}^{(r+1)}$.
    \item  With $\boldsymbol{\beta}^{(r+1)}$ and the estimated probabilities, obtain $\boldsymbol{\lambda}_{0}^{(r+1)}(\cdot)$ based on   $\tilde{\tilde{\Lambda}}_{0s}(\cdot\big|\boldsymbol{\beta}^{(r+1)})$ for $\forall s \in \mathcal{S}$.
\end{enumerate}
Repeat Steps 1 and 2 until $||\beta_s^{(r)}-\beta_s^{(r-1)}||_1/||\beta_s^{(r-1)}||_1\leq \tau^\star$ for $\forall s \in \mathcal{S}$, where $\tau^\star$ is a predetermined tolerance level. 
\end{algorithmic}
\end{algorithm}

    
\noindent The sequence limits are denoted by $\widetilde{\widetilde{\boldsymbol{\beta}}}$ and $\widetilde{\widetilde{\boldsymbol{\lambda}}}_{0}(\cdot)$, representing the estimated coefficients and baseline intensity functions obtained through our proposed approach.
\bigskip

\section{Simulation Study}\label{sec:simulation}
We conducted a simulation study to evaluate the finite-sample performance of the two approaches introduced in Sections \ref{sec:estimation_truncated} and \ref{sec:estimation_censoring}. We referred to the method based solely on zero-truncated data as Approach 1 and the one with population census information as Approach 2. We simulated a general population $\mathcal{P}$ consisting of 100,000 subjects and selected those with at least one event within a 7-year data extraction window to form the cohort $\mathcal{O}_1$. Given the heavy truncation in the MHED study, the true parameters were chosen to ensure that the cohort represented a small proportion of the population, around 0.07. The covariates considered were three indicator variables, denoted as $Z_1$, $Z_2$, and $Z_3$. The $Z_1$ followed a Bernoulli distribution with a probability of 0.5. A numerical variable $X$ was generated from a lognormal distribution, where its natural logarithm had a mean of $\log(8)$ and a variance of $\log(3)^2$. The indicators $Z_2$ and $Z_3$ were then defined as follows: $Z_{2}=I(5<X\leq 13)$, representing a medium level, and $Z_{3}=I(X>13)$, representing a high level. The stratification variable (2) was considered in the simulation study.

Three simulation scenarios were investigated in this study. In each scenario, we considered a different true model to generate event data. The true models were presented as follows.
\begin{itemize}
\item \textbf{Scenario 1}: $\lambda(a\mid \mathcal{H}_i(a), Z_i)
=\lambda_{0}\exp\{\beta_{1}Z_{i1}+\beta_{2}Z_{i2}+\beta_{3}Z_{i3}\}$ for $i \in \mathcal{P}$; that is Model (NNC).
\item \textbf{Scenario 2}: $\lambda(a\mid \mathcal{H}_i(a), Z_i)
=\lambda_{0s}\exp\{\beta_{s1}Z_{i1}+\beta_{s2}Z_{i2}+\beta_{s3}Z_{i3}\}$ for $i \in \mathcal{P}$ and $s\in\{1, 2\}$; that is Model (SSC).
\item \textbf{Scenario 3}: $\lambda(a\mid \mathcal{H}_i(a), Z_i)
=\lambda_{0s}(a)\exp\{\beta_{s1}Z_{i1}+\beta_{s2}Z_{i2}+\beta_{s3}Z_{i3}\}$ for $i \in \mathcal{P}$ and $s\in\{1, 2\}$; that is our proposed Model (\ref{eq:model}) or (SSV).
\end{itemize}
For each scenario, various estimation procedures under different models were applied to zero-truncated data and doubly censored data with all covariate information on subjects in $\mathcal{O}$. We were interested in estimating the parameters and functions under Models (NNC), (NNV), (SSC), and (1) with the generated data. Simulation results were discussed into the following subsections. All numerical results in the paper were obtained using R (Version 3.6.1). For Approach 2 under Model (\ref{eq:model}), C++ code was implemented through the R packages \textit{Rcpp} and \textit{RcppArmadillo}.

\bigskip

\noindent5.1\textit{\hspace{0.4cm}Results of Simulation Scenario 1} \label{sec:simulation_sce1}

\begin{table*}[ht!]
\centering
\caption{Estimated baselines and coefficients (standard errors) under different models using Approaches 1 (with fully known strata) and 2 with 1000 replicates and a 7-year data extraction window for Simulation Scenario 1} 

\label{Tab:simul_sc1_app1_2}
\footnotesize
\tabcolsep=4pt
\begin{subtable}{1\linewidth}
\vspace{-0.3cm}
\caption{Under models without stratification}
\begin{tabular*}{\columnwidth}{@{\extracolsep\fill}c| l| r r |r@{\extracolsep\fill}}

\hline
  
  \multicolumn{2}{c|}{$\text{Model}^{\dag}$}& \multicolumn{2}{c|}{NNC}                                                 &   \multicolumn{1}{c}{NNV}\\
 \hline
 \multirow{2}{*}{Stratum} &\multirow{2}{*}{\begin{tabular}[c]{@{}c@{}}Baseline/ Covariate\\ (True value)\end{tabular}} & \multicolumn{2}{c|}{$\widehat{\lambda}_{0s}$ or $\widehat{\beta}_s(\widehat{se}^{\star})$} &  \multicolumn{1}{c}{$\widehat{\beta}_s(\widehat{se}^{\star})$}\\
   & & $\text{Approach 1}$ & $\text{Approach 2}$ & $\text{Approach 2}$ \\ \hline
  
 \multirow{6}{*}{\begin{tabular}[c]{@{}c@{}}None\end{tabular}}& Baseline ($\lambda_{0}=0.05$)&  0.049 (0.002)           &  {0.050} (0.001)  & \multicolumn{1}{c}{\textendash}          \\  
  & Covariate & & &\\
 &\hspace{0.5 cm} $Z_1$ ($\beta_{1}=-2$)   & -1.946 (0.296)            &  {-2.000} (0.035)   &{-1.999} (0.035)        \\ 
 &\hspace{0.5 cm} $Z_2$ ($\beta_{2}=-1$)   &  {-1.013} (0.119)           & {-1.000} (0.028)  &{-1.000} (0.028)          \\ 
 &\hspace{0.5 cm} $Z_3$ ($\beta_{3}=-1.5$)   & -1.492 (0.186)            &    {-1.500} (0.034)  &{-1.500} (0.033)       \\ \hline
\end{tabular*}

\vspace{0.2cm}
\caption{Under models with stratification}
 \begin{tabular*}{\columnwidth}{@{\extracolsep\fill}c| l| r r |r@{\extracolsep\fill}}

\hline
  
  \multicolumn{2}{c|}{$\text{Model}^{\dag}$}& \multicolumn{2}{c|}{SSC}                                                 &   \multicolumn{1}{c}{SSV/ 1}\\
 \hline
 \multirow{2}{*}{Stratum} &\multirow{2}{*}{\begin{tabular}[c]{@{}c@{}}Baseline/ Covariate\\ (True value)\end{tabular}} & \multicolumn{2}{c|}{$\widehat{\lambda}_{0s}$ or $\widehat{\beta}_s(\widehat{se}^{\star})$} &  \multicolumn{1}{c}{$\widehat{\beta}_s(\widehat{se}^{\star})$}\\
   & & $\text{Approach 1}$ & $\text{Approach 2}$ & $\text{Approach 2}$ \\ \hline

\multirow{6}{*}{\begin{tabular}[c]{@{}c@{}}1\end{tabular}}&  Baseline ($\lambda_{01}=0.05$)&  $\textit{{0.027} (0.011)}$           & {0.051} (0.001)     & \multicolumn{1}{c}{\textendash}        \\ 
&Covariate& & &\\
&\hspace{0.5 cm} $Z_1$ ($\beta_{11}=-2$)   &  {-5.162} (3.244)           & {-2.006} (0.039)         & {-1.993} (0.039)   \\  
&\hspace{0.5 cm} $Z_2$ ($\beta_{12}=-1$)  &{-4.726} (3.096)            & {-1.005} (0.036)          &{-0.995} (0.036)  \\  
&\hspace{0.5 cm} $Z_3$ ($\beta_{13}=-1.5$) & {-5.333} (3.288)           &  {-1.506} (0.041)          &  {-1.494} (0.040) \\ \hline

\multirow{6}{*}{\begin{tabular}[c]{@{}c@{}}2\end{tabular}}&  Baseline ($\lambda_{02}=0.05$)&  {0.050} (0.002)           & {0.049} (0.002)     & \multicolumn{1}{c}{\textendash}        \\  
&Covariate& & &\\
&\hspace{0.5 cm} $Z_1$ ($\beta_{21}=-2$)  &  {-2.040} (0.306)           & {-2.053} (0.298)         & {-2.032} (0.294)   \\  
&\hspace{0.5 cm} $Z_2$ ($\beta_{22}=-1$)  &{-1.002} (0.118)            & {-1.009} (0.110)          &{-0.997} (0.111)  \\  
&\hspace{0.5 cm} $Z_3$ ($\beta_{23}=-1.5$) & {-1.514} (0.185)           &  {-1.523} (0.177)          &  {-1.506} (0.175) \\ \hline

\end{tabular*}

\end{subtable}%
\begin{flushleft}
\scriptsize{\noindent$\star$: The estimated standard errors are sample standard deviations (SSDs) of the 1000 estimates;\\
$\dag$: Model (NNC) is $\lambda(a\mid \mathcal{H}_i(a), Z_i)
=\lambda_{0} \exp\{\beta^{'} Z_i\}$;
Model (NNV) is $\lambda(a\mid \mathcal{H}_i(a), Z_i)
=\lambda_{0}(a)$ $ \exp\{\beta^{'} Z_i\}$;
Model (SSC) is $\lambda(a\mid \mathcal{H}_i(a), Z_i)
=\lambda_{0s} \exp\{\beta_s^{'} Z_i\}$;
Model (SSV) is $\lambda(a\mid \mathcal{H}_i(a), Z_i)
$ $= \lambda_{0s}(a)\exp\{\beta_s^{'} Z_i\}$.
}
\end{flushleft}
\end{table*}

\begin{table*}[ht!]
\centering
\caption{Estimated baselines and coefficients (standard errors) under different models using Approaches 1 (with fully known strata) and 2 with 1000 replicates  and a 7-year data extraction window for Simulation Scenario 2}
\label{Tab:simul_sc2_app1_2}
\footnotesize

\tabcolsep=4pt
\begin{subtable}{1\linewidth}
\vspace{-0.3cm}
\caption{Under models without stratification}

\begin{tabular*}{\columnwidth}{@{\extracolsep\fill}c| l| r r |r@{\extracolsep\fill}}
\hline
  
\multicolumn{2}{c|}{$\text{Model}^{\dag}$}& \multicolumn{2}{c|}{NNC}                                                 &   \multicolumn{1}{c}{NNV}\\
 \hline
 \multirow{2}{*}{Stratum} &\multirow{2}{*}{\begin{tabular}[c]{@{}c@{}}Baseline/ Covariate\end{tabular}} & \multicolumn{2}{c|}{$\widehat{\lambda}_{0s}$ or $\widehat{\beta}_s(\widehat{se}^{\star})$} &  \multicolumn{1}{c}{$\widehat{\beta}_s(\widehat{se}^{\star})$}\\
   & & $\text{Approach 1}$ & $\text{Approach 2}$ & $\text{Approach 2}$ \\ \hline
  
  \multirow{6}{*}{\begin{tabular}[c]{@{}c@{}}None\end{tabular}}& Baseline &  0.068 (0.002)           &  {0.058} (0.001)  & \multicolumn{1}{c}{\textendash}          \\  
  & Covariate& & &\\
 &\hspace{0.5 cm} $Z_1$   & -0.985 (0.111)            &  {-2.082} (0.034)   &{-2.081} (0.034)        \\ 
 &\hspace{0.5 cm} $Z_2$   &  \textit{0.457} (0.047)           & \textit{-0.580} (0.025)  &{-0.580} (0.025)          \\ 
 &\hspace{0.5 cm} $Z_3$   & -0.506 (0.088)            &    {-1.410} (0.035)  &{-1.410} (0.034)       \\ \hline
\end{tabular*}
\vspace{0.2cm}
\caption{Under models with stratification}


\begin{tabular*}{\columnwidth}{@{\extracolsep\fill}c| l| r r |r@{\extracolsep\fill}}
\hline
  
 \multicolumn{2}{c|}{$\text{Model}^{\dag}$}& \multicolumn{2}{c|}{SSC}                                                 &   \multicolumn{1}{c}{SSV/1}\\
 \hline
 \multirow{2}{*}{Stratum} &\multirow{2}{*}{\begin{tabular}[c]{@{}c@{}}Baseline/ Covariate\\ (True value)\end{tabular}} & \multicolumn{2}{c|}{$\widehat{\lambda}_{0s}$ or $\widehat{\beta}_s(\widehat{se}^{\star})$} &  \multicolumn{1}{c}{$\widehat{\beta}_s(\widehat{se}^{\star})$}\\
   & & $\text{Approach 1}$ & $\text{Approach 2}$ & $\text{Approach 2}$ \\ \hline

 \multirow{6}{*}{\begin{tabular}[c]{@{}c@{}}1\end{tabular}}&  Baseline ($\lambda_{01}=0.05$)&  $\textit{{0.027} (0.011)}$           & {0.051} (0.001)     & \multicolumn{1}{c}{\textendash}        \\  
&Covariate& & &\\
&\hspace{0.5 cm} $Z_1$ ($\beta_{11}=-2$) &  {-5.030} (3.278)           & {-2.005} (0.039)         & {-1.996} (0.038)   \\  
&\hspace{0.5 cm} $Z_2$ ($\beta_{12}=-1$)  &{-4.751} (3.180)            & {-1.004} (0.033)          &{-0.995} (0.033)  \\  
&\hspace{0.5 cm} $Z_3$ ($\beta_{13}=-1.5$) & {-5.427} (3.341)           &  {-1.506} (0.042)          &  {-1.496} (0.041) \\ \hline

  \multirow{6}{*}{\begin{tabular}[c]{@{}c@{}}2\end{tabular}}&  Baseline ($\lambda_{02}=0.07$) &  {0.070} (0.002)           & {0.069} (0.002)     & \multicolumn{1}{c}{\textendash}        \\  
&Covariate& & &\\
&\hspace{0.5 cm} $Z_1$ ($\beta_{21}=-1$)  &  {-1.007} (0.111)           & {-1.012} (0.107)         & {-1.005} (0.106)   \\  
&\hspace{0.5 cm} $Z_2$ ($\beta_{22}=0.5$)  &{0.498} (0.048)            & {0.503} (0.046)          &{0.485} (0.046)  \\  
&\hspace{0.5 cm} $Z_3$ ($\beta_{23}=-0.5$) & {-0.502} (0.087)           &  {-0.504} (0.085)          &  {-0.505} (0.084) \\ \hline

\end{tabular*}
\end{subtable}%
\begin{flushleft}
    \scriptsize{\noindent$\star$: The estimated standard errors are sample standard deviations (SSDs) of the 1000 estimates;\\
    $\dag$: Model (NNC) is $\lambda(a\mid \mathcal{H}_i(a), Z_i)
=\lambda_{0} \exp\{\beta^{'} Z_i\}$;
Model (NNV) is $\lambda(a\mid \mathcal{H}_i(a), Z_i)
=\lambda_{0}(a)$ $ \exp\{\beta^{'} Z_i\}$;
Model (SSC) is $\lambda(a\mid \mathcal{H}_i(a), Z_i)
=\lambda_{0s} \exp\{\beta_s^{'} Z_i\}$;
Model (SSV) is $\lambda(a\mid \mathcal{H}_i(a), Z_i)
$ $= \lambda_{0s}(a)\exp\{\beta_s^{'} Z_i\}$.
}
\end{flushleft}
\end{table*}

\noindent Table \ref{Tab:simul_sc1_app1_2} shows the estimation results for Approaches 1 and 2 under different models. Due to the computational complexity of Approach 1, we only considered the case where strata were fully known. Under Model (NNC), both approaches produce estimates close to the true values while the estimated standard errors (SEs) in Approach 1 are much larger than those in Approach 2. Under Model (SSC), Approach 2 performs notably better, particularly in stratum 1. For Approach 1, the estimated baseline in stratum 1 (\textit{italicized}) is significantly different from the true parameter and the estimated coefficients are not significantly different from 0 due to the large SEs in the same stratum. The large SEs reflect large uncertainty before the first observed event among subjects born before the study, as the stratum before this event is unknown, and all subjects must be in stratum 2 after experiencing an event. Given the poor estimation results from Approach 1 with fully known strata under Model (SSC), there is no need to apply Approach 1 with partially known strata since the results would be even worse with less information. Moreover, Approach 2 provides consistent coefficient estimates under models with constant and arbitrary baseline intensity functions.

Table S.1 in Supplementary Material C presents the estimation results for doubly censored data with fully and partially known strata, representing the best possible performance of Approach 2. A comparison between Tables \ref{Tab:simul_sc1_app1_2} and S.1 reveals that Approach 2 provides competitive estimation results compared to those obtained using complete covariate and censoring information, which is also supported by Figure S.2 in Supplementary Material B. Additionally, the estimation results with and without fully known strata are quite similar, indicating a good approximation of the probability of being in stratum 1 or 2 under different models. However, if cohort $\mathcal{O}_1$ is mistakenly treated as a random sample of the general population $\mathcal{P}$, the estimation results become quite biased and misleading (see Table S.2 in Supplementary Material C).

\bigskip

\noindent5.2\textit{\hspace{0.4cm}Results of Simulation Scenario 2} \label{sec:simulation_sce2}

\noindent In Scenario 2, similar estimation tables and plots were provided. Since many of the discussions are similar to those in Scenario 1, we will not repeat them in the following simulation results. In Table \ref{Tab:simul_sc2_app1_2}, the \textit{italicized} estimates under Model (NNC) are inconsistent between Approaches 1 and 2. Specifically, the estimated coefficient for $Z_2$ is significantly positive in Approach 1, while it is significantly negative in Approach 2. This difference suggests that the two approaches could lead to opposite conclusions when stratification is mistakenly omitted from the model and the true parameters in different strata have opposite signs. Moreover, the estimates in Approach 2 are close to those obtained from the doubly censored data, regardless of whether the model is mis-specified.

\bigskip

\noindent5.3\textit{\hspace{0.4cm}Results of Simulation Scenario 3} \label{sec:simulation_sce3}

\noindent In Scenario 3, the true model contains stepwise baseline intensity functions in both strata. The first two columns of Table \ref{Tab:simul_sc3_app1_2} present results for the case where the model is mis-specified. In this case, our proposed approach, Approach 2, still provides accurate coefficient estimates, demonstrating the robustness of the associated estimators. In contrast, Approach 1 fails to produce a good estimation of coefficients in stratum 1; the estimated coefficients for $Z_1$ and $Z_3$ (italicized) are significantly different from the true values.  

\bigskip

To summarize, we have verified that the estimators resulting from our proposed method (Approach 2) are both consistent and robust in the three simulation settings. Our approach utilizes much less information but has competitive estimation performance compared to the approaches using complete covariate 

\newpage
\begin{table*}[ht!]
\centering
\caption{\small Estimated baselines and coefficients (standard errors) under different models using Approaches 1 (with fully known strata) and 2 with 1000 replicates  and a 7-year data extraction window for Simulation Scenario 3} 
\scriptsize
\label{Tab:simul_sc3_app1_2}
\tabcolsep=4pt
\begin{subtable}{1\linewidth}
\vspace{-0.3cm}
\caption{Under models without stratification}

\begin{tabular*}{\columnwidth}{@{\extracolsep\fill}c| l| r r |r@{\extracolsep\fill}}
\hline
  
\multicolumn{2}{c|}{$\text{Model}^{\star\star}$}& \multicolumn{2}{c|}{NNC}                                                 &   \multicolumn{1}{c}{NNV}\\
 \hline
 \multirow{2}{*}{Stratum} &\multirow{2}{*}{\begin{tabular}[c]{@{}c@{}}Baseline/ Covariate\end{tabular}} & \multicolumn{2}{c|}{$\widehat{\lambda}_{0s}$ or $\widehat{\beta}_s(\widehat{se}^{\star})$} &  \multicolumn{1}{c}{$\widehat{\beta}_s(\widehat{se}^{\star})$}\\
   & & $\text{Approach 1}$ & $\text{Approach 2}$ & $\text{Approach 2}$ \\ \hline
  
  \multirow{6}{*}{\begin{tabular}[c]{@{}c@{}}None\end{tabular}}& Baseline &  0.057 (0.002)           &  {0.046} (0.001)  & \multicolumn{1}{c}{\textendash}          \\  
  & Covariate& & &\\
 &\hspace{0.5 cm} $Z_1$   & -1.006 (0.135)            &  {-2.064} (0.038)   &{-2.064} (0.038)        \\ 
 &\hspace{0.5 cm} $Z_2$   &  \textit{0.490} (0.060)           & \textit{-0.637} (0.030)  &{-0.637} (0.030)          \\ 
 &\hspace{0.5 cm} $Z_3$   & -0.501 (0.114)            &    {-1.419} (0.037)  &{-1.420} (0.036)       \\ \hline
\end{tabular*}
\vspace{0.2cm}
\caption{Under models with stratification}


\begin{tabular*}{\columnwidth}{@{\extracolsep\fill}c| l| r r |r@{\extracolsep\fill}}
\hline
  
 \multicolumn{2}{c|}{$\text{Model}^{\star\star}$}& \multicolumn{2}{c|}{SSC}                                                 &   \multicolumn{1}{c}{SSV/ 1}\\
 \hline
 \multirow{2}{*}{Stratum} &\multirow{2}{*}{\begin{tabular}[c]{@{}c@{}}Baseline/ Covariate\\ (True value)\end{tabular}} & \multicolumn{2}{c|}{$\widehat{\lambda}_{0s}$ or $\widehat{\beta}_s(\widehat{se}^{\star})$} &  \multicolumn{1}{c}{$\widehat{\beta}_s(\widehat{se}^{\star})$}\\
   & & $\text{Approach 1}$ & $\text{Approach 2}$ & $\text{Approach 2}$ \\ \hline

 \multirow{6}{*}{\begin{tabular}[c]{@{}c@{}}1\end{tabular}}&  $\text{Baseline}^{\dag}$ &  {0.000} (0.001)           & {0.036} (0.001)     & \multicolumn{1}{c}{\textendash}        \\  
&Covariate& & &\\
&\hspace{0.5 cm} $Z_1$ ($\beta_{11}=-2$)  &  \textit{0.696} (0.740)           & {-1.932} (0.045)         & {-2.006} (0.043)   \\  
&\hspace{0.5 cm} $Z_2$ ($\beta_{12}=-1$)  &{0.234} (0.729)            & {-0.986} (0.038)          &{-0.994} (0.036)  \\  
&\hspace{0.5 cm} $Z_3$ ($\beta_{13}=-1.5$) & \textit{0.008} (0.700)           &  {-1.449} (0.044)          &  {-1.503} (0.043) \\ \hline

  \multirow{6}{*}{\begin{tabular}[c]{@{}c@{}}2\end{tabular}}&  $\text{Baseline}^{\ddag}$ &  {0.054} (0.002)           & {0.075} (0.003)     & \multicolumn{1}{c}{\textendash}        \\  
&Covariate& & &\\
&\hspace{0.5 cm} $Z_1$ ($\beta_{21}=-1$)  &  {-1.064} (0.149)           & {-0.906} (0.118)         & {-0.997} (0.128)   \\  
&\hspace{0.5 cm} $Z_2$ ($\beta_{22}=0.5$)  &{0.525} (0.061)            & {0.454} (0.056)          &{0.484} (0.059)  \\  
&\hspace{0.5 cm} $Z_3$ ($\beta_{23}=-0.5$) & {-0.538} (0.123)           &  {-0.445} (0.101)          &  {-0.503} (0.109) \\ \hline

\end{tabular*}
\end{subtable}%
\begin{flushleft}
    \scriptsize{\noindent$\star$: The estimated standard errors are sample standard deviations (SSDs) of the 1000 estimates;\\
    $\star\star$: Model (NNC) is $\lambda(a\mid \mathcal{H}_i(a), Z_i)
=\lambda_{0} \exp\{\beta^{'} Z_i\}$;
Model (NNV) is $\lambda(a\mid \mathcal{H}_i(a), Z_i)
=\lambda_{0}(a)$ $ \exp\{\beta^{'} Z_i\}$;
Model (SSC) is $\lambda(a\mid \mathcal{H}_i(a), Z_i)
=\lambda_{0s} \exp\{\beta_s^{'} Z_i\}$;
Model (SSV) is $\lambda(a\mid \mathcal{H}_i(a), Z_i)
$ $= \lambda_{0s}(a)\exp\{\beta_s^{'} Z_i\}$;\\
$\dag$: The true function $\lambda_{01}(a)=0.03$ for $0 \leq a \leq 11$, $\lambda_{01}(a)=0.06$ for $11 < a \leq 18$; \\
$\ddag$: The true function $\lambda_{02}(a)=0.04$ for $0 \leq a \leq 11$, $\lambda_{02}(a)=0.08$ for $11 < a \leq 18$.
}
\end{flushleft}
\end{table*}

\noindent and observation period information. The estimated SEs are only slightly larger than those based on more information.

\section{Analysis of the MHED Dataset}\label{sec:real_data}

The MHED records of Alberta residents under 18 years old during April 1, 2010 to March 31, 2017 were extracted from the National Ambulatory Care Reporting System (NACRS; \citealp{NACRS}) developed by the Canadian Institute for Health Information (CIHI). The MHED data were linked to an annual Cumulative Registry File (CRF) to obtain the demographic and geographic information on the subjects with MHED visits. In our study period, there were 33,299 Alberta children and youth making 58,166 MHED presentations in total. About 67.1\% of subjects had one MHED visit. The percentages for 2 and 3 visits were 17.2 and 7.2, respectively. Less than 0.8\% of subjects made 10 or more MHED presentations in Alberta during the 7 years. About 88.7\% of MHED visits were made by subjects over 11 years old.

\begin{table}[ht!]
\centering
\caption{Overview of the MHED dataset (2010-2017)  }

\small
\label{Tab:PMHC_summary}
\begin{tabular}{c|r|rr|rrr}
\hline
\multirow{2}{*}{} & \multirow{2}{*}{Total} &\multicolumn{2}{c|}{Sex}   & \multicolumn{3}{c}{Region of residence}  \\ \cline{3-7} 
  &     &\multicolumn{1}{c}{Female}     & \multicolumn{1}{c|}{Male}   & \multicolumn{1}{c}{Edmonton} & \multicolumn{1}{c}{Calgary} &  \multicolumn{1}{r}{Others}\\ \hline

Subjects & 33299    
& 19210 & 14089                   
 & 8365 &  11579  & 13355\\ \hline
Visits & 58166                   
&35937 & 22229                  
 & 15135 & 20210  & 22821\\ \hline
\end{tabular}
\end{table}

We consider two covariates: \textit{sex} (female vs male) and \textit{Region of residence} (Calgary, Edmonton, and others). Since the two covariates are relatively stable across time, we treat them as time-independent variables in the following analyses. If the covariates changed for a few subjects during the study period, we used the information from their first observed MHED visits. Table \ref{Tab:PMHC_summary} shows that more females and residents in other regions visited EDs for mental health conditions. When we closely look at the first, second, third and subsequent observed MHED visits in Table S.7 in Supplementary Material C, we observe a similar pattern that females, residents in other regions, and subjects over 11 contributed most of MHED visits. Compared with the population census information in 2017 (see Table \ref{Tab:census2017}), there were more males and Calgary residents in the population under 18 years old, which suggests that the MHED cohort is not representative to Alberta population under 18 years old. 

\begin{table}
\centering
\caption{Population census information of Alberta residents under 18 years old in 2017}
\label{Tab:census2017}
\small

\begin{tabular}{r|rrr|r}
  \hline
Sex  & Edmonton & Calgary  & Others &Total\\ 
  \hline
Female  & 145499 & 172982  & 145525 &464006\\ 
  Male & 151094 & 181462  & 151576  &484132\\ 
    \hline
  Total  & 296593 & 354444  &297101 & 948138\\
   \hline
\end{tabular}
\end{table}

We applied the estimation procedures described in Sections \ref{sec:truncation_P_avail_S} and \ref{sec:censoring_known_S} to the MHED dataset under Models (NNC), (NNV), (SSC), and (1) or (SSV). We considered the stratification variable (\ref{eq:stratification_variable}) in the real data analysis. Since the log-likelihood function under Model (SSC) is complicated already for zero-truncated recurrent events only (see Appendix \ref{sec:Appendix_Estimation_details}2), we did not conduct the analysis under models with unspecified baselines and with partially known strata for zero-truncated data only. Table \ref{Tab:real_S} shows the estimated baselines and coefficients with the estimated SEs under Model (SSC) and Model (\ref{eq:model}) using Approach 
2. The \textbf{bolded} estimates are significantly different from zero. The estimated coefficients under different models are consistent except the one for Calgary. Under Model (SSC), the estimated coefficient for Calgary is not significantly different from zero in stratum 2, whereas it is significantly positive under Model (\ref{eq:model}).
Moreover, the opposite signs of the estimated coefficients for Edmonton in the two strata under both models suggests that, compared with residents in other regions, children and youth in Edmonton had a lower risk of experiencing their first MHED visit, given a fixed sex. However, once they made their first visit, they were more likely to have subsequent MHED visits. Other estimates can be interpreted in a similar way. 

\begin{table}[ht!]
\centering
\caption{Estimated baselines and coefficients (standard errors) under models with stratification in $\text{Approach 2}$ for the MHED study} 
\small
\label{Tab:real_S}
%

\begin{tabular}{c| l| r |r} 

\hline
  
  \multicolumn{2}{c|}{$\text{Model}^{\dag}$}& \multicolumn{1}{c|}{SSC}                                                 &   \multicolumn{1}{c}{SSV/ 1}\\
 \hline
 \multirow{1}{*}{Stratum} &\multirow{1}{*}{Baseline/ Covariate} & \multicolumn{1}{c|}{$\widehat{\lambda}_{0s}$ or $\widehat{\beta}_s(\widehat{se}^{\star})$} &  \multicolumn{1}{c}{$\widehat{\beta}_s(\widehat{se}^{\star})$}\\ \hline

  
 \multirow{4}{*}{1}& Baseline            &  \textbf{0.004} (0.000)  & \multicolumn{1}{c}{\textendash}          \\  
 & Sex (vs Female)& & \\
 &\hspace{0.5 cm} Male              &  \textbf{-0.346} (0.013)   &\textbf{-0.379} (0.013)        \\ 
  & Region (vs Others) &  &\\
 &\hspace{0.5 cm} Edmonton              & \textbf{-0.387} (0.015)  &\textbf{-0.380} (0.015)          \\ 
 &\hspace{0.5 cm} Calgary             &    \textbf{-0.247} (0.014)  &\textbf{-0.277} (0.015)       \\ \hline
 
\multirow{4}{*}{2}&  Baseline            & \textbf{0.311} (0.004)     & \multicolumn{1}{c}{\textendash}        \\  
&Sex (vs Female)& & \\
&\hspace{0.5 cm} Male            & \textbf{-0.269} (0.017)         & \textbf{-0.383} (0.029)   \\  
 &Region (vs Others) & & \\
&\hspace{0.5 cm} Edmonton            & \textbf{0.094} (0.022)          &\textbf{0.140} (0.037)  \\  
&\hspace{0.5 cm} Calgary            &  0.030 (0.020)          &  \textbf{0.151} (0.030) \\ \hline
\end{tabular}
\begin{flushleft}
    \scriptsize{\noindent$\star$: The estimated standard errors are calculated based on 1000 bootstrap samples;\\
        $\dag$: Model (SSC) is $\lambda(a\mid \mathcal{H}_i(a), Z_i)
=\lambda_{0s} \exp\{\beta_s^{'} Z_i\}$;
Model (SSV) is $\lambda(a\mid \mathcal{H}_i(a), Z_i)
= \lambda_{0s}(a)$ $\exp\{\beta_s^{'} Z_i\}$.
}
\end{flushleft}
\end{table}

\begin{table}[ht!]
\centering
\caption{Estimated baselines and coefficients (standard errors) under models without stratification in different approaches for the MHED study} 
\small
\label{Tab:real_noS}
\begin{tabular}{l| r r |r}

\hline
  
   \multicolumn{1}{c|}{$\text{Model}^{\dag}$}& \multicolumn{2}{c|}{NNC}                                       &   \multicolumn{1}{c}{NNV}\\
 \hline
 \multirow{2}{*}{Baseline/ Covariate}  & \multicolumn{2}{c|}{$\widehat{\lambda}_0$ or $\widehat{\beta}(\widehat{se}^{\star})$} &  \multicolumn{1}{c}{$\widehat{\beta}(\widehat{se}^{\star})$}\\
   &  $\text{Approach 1}$ & $\text{Approach 2}$ & $\text{Approach 2}$\\ \hline
  Baseline&  \textbf{0.240} $(0.005)$    & \textbf{0.014} (0.000)        & \multicolumn{1}{c}{\textendash}     \\ 
 Sex (vs Female)& & &\\
  \hspace{0.5 cm} Male& \textbf{-0.327} (0.022)  & \textbf{-0.530} (0.016)  & \textbf{-0.536} (0.016)          \\ 
 Region (vs Others) & & &\\
 \hspace{0.5 cm} Edmonton& \textbf{0.113} (0.028)             & \textbf{-0.321} (0.020) & \textbf{-0.301} (0.021)          \\ 
  \hspace{0.5 cm} Calgary & 0.015 (0.026)            & \textbf{-0.231} (0.018)       & \textbf{-0.196} (0.018)                \\ \hline
\end{tabular}
\begin{flushleft}
    \scriptsize{\noindent$\star$: The estimated standard errors are calculated based on 1000 bootstrap samples;\\
        $\dag$: Model (NNC) is $\lambda(a\mid \mathcal{H}_i(a), Z_i)
=\lambda_{0} \exp\{\beta^{'} Z_i\}$;
Model (NNV) is $\lambda(a\mid \mathcal{H}_i(a), Z_i)
=\lambda_{0}(a)$ $ \exp\{\beta^{'} Z_i\}$.
    \footnotesize}
\end{flushleft}
\end{table}

\begin{figure}[!ht]
\centering
\includegraphics[width=0.8\textwidth]{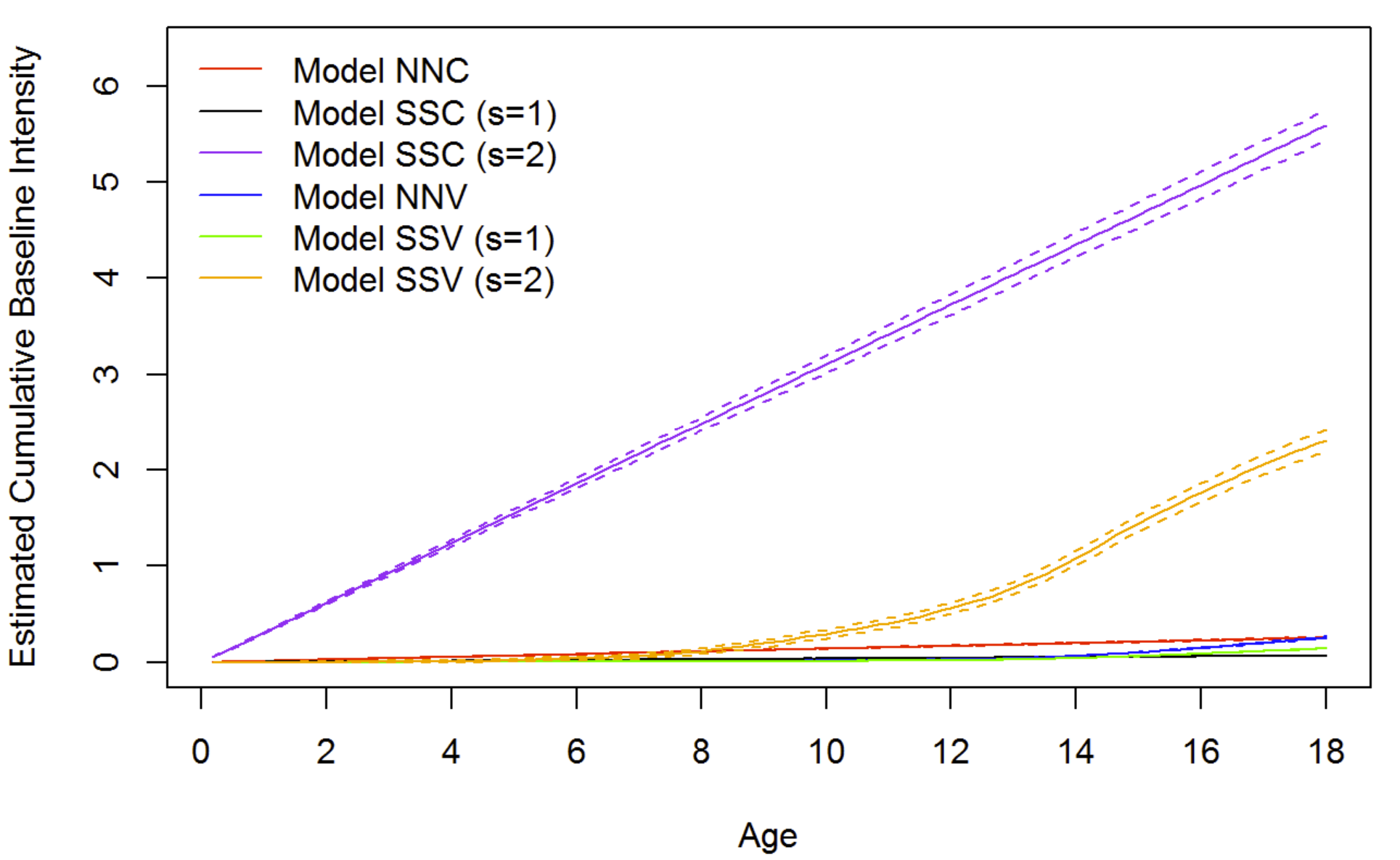}
\caption{\label{Figure:real_exa_cum_baseline}Estimated cumulative baseline intensity functions with 95\% pointwise confidence intervals for the MHED study in Approach 2.\\ \footnotesize{\noindent Note: Model (SSV) in the plots is Model (\ref{eq:model}).}}
\end{figure}

We also considered a special case where there was no stratum. Then, Model (\ref{eq:model}) becomes the Andersen-Gill (AG) model \citep{AGmodel}. Table \ref{Tab:real_noS} presents the estimation results under models without stratification using Approaches 1 and 2. Under Model (NNC), the two approaches give us inconsistent results for Edmonton and Calgary. The results of Approach 1 suggest that, for a given sex, residents in Edmonton were more likely to have MHED visits. In contrast, the results of Approach 2 show that residents in Edmonton and Calgary had a lower risk of making MHED visits compared to those in other regions. 
Comparison of the results in Tables \ref{Tab:real_S} and \ref{Tab:real_noS} shows notable differences in the estimates across strata, which highlights the necessity of including the stratification variable in the MHED study.


Figure \ref{Figure:real_exa_cum_baseline} shows the estimated cumulative baseline intensity functions under different models with 95\% pointwise confidence intervals in Approach 2. In stratum 1, the estimated function for an arbitrary baseline is close to that for a constant baseline, while they diverge in stratum 2. This difference is also reflected in Table \ref{Tab:real_S}, where the estimated coefficients in stratum 2 have relatively large differences between different models.

\section{Conclusion}\label{sec:conclusion}

We present two estimation approaches for analyzing zero-truncated recurrent event data under an innovative intensity-based model. This model depends on the event history only through a stratification variable. With the stratification variable, only a summary of the past events, such as whether a subject has experienced their first event, is needed to achieve our research objective. Our estimation procedures address the challenges of analyzing zero-truncated recurrent event data with partially known strata. In our proposed method (Approach 2), we adapt \cite*{Yi_Hu_Rosychuk2020}'s approach of integrating zero-truncated data with population census information. Compared to the MLE (Approach 1), Approach 2 is computationally more efficient and yields consistent and robust estimates. In the MHED study, we find that the covariate effects vary across strata, providing insights into how the previous visits influence the occurrence of the later visits. Based on the estimation results, we can answer questions related to the rate of having the first MHED visit, the rate of having subsequent visits given the first one, and the marginal rate of visits for certain subjects at a given age. Additionally, we can compare the rate differences across different subjects with a given summary of their event histories. 

The model assumptions may be problematic and merit further discussions. First, we assume independence between subjects, which might not hold in reality. Children and youth with mental health conditions can cluster within communities. For instance, individuals from low-income areas may have less access to alternative mental health services and visit EDs more frequently for mental health crisis. In our study, we focus on large regions like Edmonton, Calgary, and other regions, where correlation is less of a concern. However, in studies conducted in smaller regions, researchers should account for correlations based on geographic locations. The assumption that the data extraction window is independent of the recurrent events is reasonable in our case because the window was selected without looking at the data. If researchers focus on specific time periods with potential changes in the number of events, such as MHED visits during the COVID-19 pandemic, this assumption could be violated. Lastly, we assume that birthdates are independent of the counting process, conditional on covariates. This could also be problematic since different generations may have distinct event patterns, as noted by \cite*{Yi_Hu_Rosychuk2020}.

\cite{nirmalkanna2024analysis} proposed intensity-based models including carryover effects and trends due to previous events, which is related to our Model (SNC) and is further discussed. Their model (3) is expressed as 
\begin{equation*}
\lambda_i(t\mid \mathcal{H}_i(t))
=\lambda_{0i}(t) \exp\{\gamma N_i(t-)+\beta Z_i(t)+ \xi'x_i(t)\},
\end{equation*}
where $\lambda_{0i}(t)=\alpha$ for $i=1,\cdots, m$, $\alpha>0$, and $t>0$. In this model, $\gamma$ represents a monotonic trend associated with the number of previous events $N_i(t-)$ and $\beta$ denotes a carryover effect. The indicator $Z_i(t)=I(N_i(t-)>0)I(B_i(t)\leq \Delta)$ captures whether a previous event occurred and whether the elapsed time since the last event $B_i(t)$ falls within a predefined duration of the carryover effect $\Delta$. The $\xi$ corresponds to the effects of external covariates $x_i(t)$. An equivalent representation of Model (SNC) in our study is written as 
\begin{equation}
\lambda(a\mid \mathcal{H}_i(a), Z_i)
=\lambda_{0} \exp\{\alpha I(S_i(a)=s)+\beta^{'} Z_i\}.
\label{eq:model_SNC_equiv}
\end{equation}
With the stratification variable (\ref{eq:stratification_variable}), $I(S_i(a)=2)\equiv I(N_i(a-)>0)$. Then, our Model (\ref{eq:model_SNC_equiv}) is equivalent to their Model (3) when the trend effect is excluded, $\Delta$ is large, and the covariates are time-independent. In terms of estimation procedures, \cite{nirmalkanna2024analysis} employed maximum likelihood estimation to estimate the parameters, $(\gamma,\beta,\xi')'$. However, this method is not suitable in our study due to the nature of zero-truncated recurrent event data. 

Some practical issues remain for future investigation. Early studies of the PMHC program showed that covariate effects could vary over time under marginal models \citep*{Hu_Rosychuk2016,Yi_Hu_Rosychuk2020}. This motivates us to explore the estimation of time-varying coefficients under our proposed model for future research. Furthermore, we integrate the MHED data with Alberta census information to make an inference on a general population $\mathcal{P}$. This approach replies on two assumptions: that Alberta residents under 18 can reasonably be treated as a random sample of the general population, and that event patterns are similar over time. However, the second assumption may not hold since different generations could have varying event patterns. To address this, we could apply our analyses to MHED datasets extracted from different time windows to examine changes in event patterns over time. Additionally, we could extend our analyses to multi-type recurrent events, such as investigating MHED visits related to specific mental health disorders or different severity levels. It would also be valuable to explore a multi-state representation of the recurrent event process.

\newpage
\appendix

\section*{Appendices}

\section{Details for Estimation Procedures} \label{sec:Appendix_Estimation_details}
\normalsize
\subsubsection*{A1. The expression of the induced model with the stratification variable (\ref{eq:stratification_variable})}

\noindent The induced model is written as
\begin{align*}
    \lambda^{\star}(a\mid \mathcal{H}_i(a), Z_i)=\lambda(a\mid \mathcal{H}_i(a), Z_i)\frac{P\bigl(N^{\star}_i\geq1|dN_i(a)=1, \mathcal{H}_i(a), Z_i\bigr)}{P\bigl(N^{\star}_i\geq1|\mathcal{H}_i(a), Z_i\bigr)}.
\end{align*}
Let $R(a, Z_i)=P\bigl(N^{\star}_i\geq1|\mathcal{H}_i(a), Z_i\bigr)/P\bigl(N^{\star}_i\geq1|dN_i(a)=1, \mathcal{H}_i(a), Z_i\bigr)$, then $ \lambda^{\star}(a\mid \mathcal{H}_i(a), Z_i)=\lambda(a\mid \mathcal{H}_i(a), $ $Z_i)/R(a, Z_i)$. Given stratification variable (\ref{eq:stratification_variable}), 
\begin{align*}
\begin{split}
  &R(a, Z_i)\\
&=\begin{cases}\frac{1-\exp\Bigl\{-\int_{a}^{C_{R_i}}\lambda_{01}(u) e^{\beta_1^{'}Z_i } du \Bigr\}-
\left[1-\exp\Bigl\{-\int_{a}^{C_{L_i}}\lambda_{01}(u)e^{\beta_1^{'}Z_i } du\Bigr\}\right]\exp\Bigl\{-\int_{C_{L_i}}^{C_{R_i}}\lambda_{02}(u)e^{\beta_2^{'}Z_i }du \Bigr\}}{1-\exp\bigl\{-\int_{C_{L_i}}^{C_{R_i}}\lambda_{02}(u)e^{\beta_2^{'}Z_i }du \bigr\}}, \\& \hspace{-2.2cm} 0<a \leq C_{L_i}\\
1-\exp\bigl\{-\int_{a}^{C_{R_i}}\lambda_{01}(u) e^{\beta_1^{'}Z_i }du\bigr\}, & \hspace{-3.5cm} C_{L_i}<a \leq C_{R_i}<A^\star\end{cases}
\end{split}
\end{align*}
\normalsize
for $s=1$;
\begin{align*}
\begin{split}
&R(a, Z_i)\\
&=\begin{cases}1-\exp\bigl\{-\int_{a}^{C_{R_i}}\lambda_{02}(u) e^{\beta_2^{'}Z_i } du\bigr\} & C_{L_i}<a \leq C_{R_i},~ N_i(a-)-N_i(C_{L_i})=0\\ 
1& \text{otherwise}\end{cases}
\end{split}
\end{align*}
for $s=2$ and $i\in\mathcal{P}_1$.
\bigskip

\bigskip

\subsubsection*{A2. The log-likelihood function for zero-truncated data with the fully known stratification variable and $\lambda_{0s}(a)=\lambda_{0s}$}

\noindent Let $Y_i^{(c)}(a)= I\big(a \in (C_{Li}, C_{Ri}]\big)$, $Y_i^{(s)}(a)=I\big(a: S_i(a)=s\big)$, and $Y_i^\star(a)=I\big(a: N_i(a-)-N_i(C_{L_i})=0\big)$. Given the stratification variable (\ref{eq:stratification_variable}), the log-likelihood function of $\boldsymbol{\alpha}$ is

\scriptsize
\begin{align*}
\begin{split}
\ell^\star(\boldsymbol{\alpha}|\boldsymbol{S}(\cdot),\mathcal{Q}_{1})
   &=\sum_{i\in \mathcal{O}_1}^{} \Biggl\{ \sum_{s=1}^{2}\biggl(\int_{0}^{A^\star} Y_i^{(s)}(a)Y_i^{(c)}(a)Y_i^\star(a)\left[\alpha_s^{'}Z_i^\star-
    \log\left(1-\exp\bigl\{-e^{\alpha_s^{'}Z_i^\star}(C_{R_i}-a)\bigr\}\right)\right] dN_i(a)  \\
     &\quad -\int_{0}^{A^\star}Y_i^{(s)}(a)Y_i^{(c)}(a)Y_i^\star(a)\frac{e^{\alpha_s^{'}Z_i^\star}}{1-\exp\bigl\{-e^{\alpha_s^{'}Z_i^\star}(C_{R_i}-a)\bigr\}} da \biggr)  \\
     &\quad +\int_{0}^{A^\star} Y_i^{(2)}(a)Y_i^{(c)}(a)\bigl[1-Y_i^\star(a)\bigr]\alpha_2^{'}Z_i^\star dN_i(a)
     -\int_{0}^{A^\star}Y_i^{(2)}(a)Y_i^{(c)}(a)\bigl[1-Y_i^\star(a)\bigr]e^{\alpha_2^{'}Z_i^\star}da \Biggr\},
\end{split}
\end{align*}
\normalsize
where $\alpha_s=(\log\lambda_{0s},\beta_s^{'})^{'}$, $Z^\star_i=(1, Z_i^{'})^{'}$, and $\boldsymbol{\alpha}=\{\alpha_{s}:s\in\{1,2\}\}$.
\bigskip

\subsubsection*{A3. The calculation of $L(\boldsymbol{\alpha}|\mathcal{Q}_1 )$ with the stratification variable (\ref{eq:stratification_variable})}

\noindent The likelihood function based on zero-truncated data with the partially known stratification variable is
\scriptsize
\begin{align*}
\begin{split}
L(\boldsymbol{\alpha}|\mathcal{Q}_1 )&=\prod_{i\in\mathcal{O}_1}\bigg\{ \sum_{s=1}^2\Big[\lambda^{\star}_s\big(a_{i1}\mid \mathcal{H}_i(a_{i1}), Z_i\big)exp\big\{-\int_{C_{L_i}}^{a_{i1}} \lambda^{\star}_s(a\mid \mathcal{H}_i(a), Z_i)da\big\}P\big(Y_i^{(s)}(C_{L_i})=1|Z_i\big)\Big]\\
&\Big[\prod_{j=2}^{N^\star_i}\lambda^{\star}_2\big(a_{ij}\mid \mathcal{H}_i(a_{ij}), Z_i\big)\Big]exp\Big\{-\int_{a_{i1}}^{C_{R_i}} \lambda^{\star}_2\big(a\mid \mathcal{H}_i(a), Z_i\big)da\Big\}\bigg\},
\end{split}
\end{align*}
\normalsize
where $\lambda^{\star}_s(a\mid \mathcal{H}_i(a), Z_i)=Y_i^{(s)}(a)\lambda^{\star}(a\mid \mathcal{H}_i(a), Z_i)$ and
\begin{align*}
    &P(Y_i^{(s)}(C_{L_i})=1|Z_i)\\
    &=\begin{cases}exp\Big\{-\int_{0}^{C_{L_i}} \lambda^{\star}_1\big(a\mid \mathcal{H}_i(a), Z_i\big)da\Big\}, &\text{ }s=1 \\1-exp\Big\{-\int_{0}^{C_{L_i}} \lambda^{\star}_1\big(a\mid \mathcal{H}_i(a), Z_i\big)da\Big\}, &\text{ }s=2\end{cases}.
\end{align*}
When $N^\star_i=1$ for $i\in\mathcal{O}_1$, we let $\prod_{j=2}^{N^\star_i}\lambda^{\star}_2\big(a_{ij}\mid \mathcal{H}_i(a_{ij}), Z_i\big)=1$.

\bigskip

\subsubsection*{A4. The calculation of $P\bigl(Y_i^{(s)}(a)=1|\mathcal{Q}_{1i}\bigr)$ with the stratification variable (\ref{eq:stratification_variable})}

\noindent For $i\in \mathcal{O}_1$ and $a\in(C_{L_i},C_{R_i}]$, the probability
\begin{align*}
    &P\bigl(Y_i^{(1)}(a)=1|\mathcal{Q}_{1i}\bigr)\\
    &=P(N_i(a-)=0|\mathcal{Q}_{1i})\\
    &=\begin{cases}0, &\exists\text{ } a_{ij}\in(C_{L_i},C_{R_i}],~a_{ij}<a\\P\big(N_i(C_{L_i})=0,N_i(a-)-N_i(C_{L_i})=0|\mathcal{Q}_{1i}\big), &\text{otherwise}\end{cases},
\end{align*}
where $a_{ij}$ is the age of subject $i$ at the $j$-th observed event time within $(C_{L_i},C_{R_i}]$ for $j=1,\cdots,N_i^{\star}$. 

The probability
    $P\big(N_i(C_{L_i})=0,N_i(a-)-N_i(C_{L_i})=0|\mathcal{Q}_{1i}\big)
    $ $=P\big(N_i(C_{L_i})=0,N_i(a-)-N_i(C_{L_i})=0, \mathcal{Q}_{1i}|Z_i\big)/P\big(\mathcal{Q}_{1i}|Z_i\big)$. The numerator is 
\begin{align*}
    &P\big(N_i(C_{L_i})=0,N_i(a-)-N_i(C_{L_i})=0,\mathcal{Q}_{1i}|Z_i\big)\\
    &=P\big(N_i(C_{L_i})=0|Z_i)P\big(N_i(a-)-N_i(C_{L_i})=0,\mathcal{Q}_{1i}|N_i(C_{L_i})=0,Z_i\big)\\
&=\lambda_{01}(a_{i1})e^{\beta_1^{'}Z_i}\exp\big\{-\int_{0}^{a_{i1}}\lambda_{01}(u)e^{\beta_1^{'}Z_i}du\big\}\\
&\hspace{0.5cm}\times\Big[\prod_{j=2}^{N_i^\star}\big\{\lambda_{02}(a_{ij})e^{\beta_2^{'}Z_i}\big\}\Big]\exp\big\{-\int_{a_{i1}}^{C_{R_i}}\lambda_{02}(u)e^{\beta_2^{'}Z_i}du\big\}.  
\end{align*}
When $N_i^\star=1$ for $i\in\mathcal{O}_1$, $\prod_{j=2}^{N_i^\star}\big\{\lambda_{02}(a_{ij})e^{\beta_2^{'}Z_i}\big\}=1$ in the formula above. Similarly, the denominator is $P\big(\mathcal{Q}_{1i}|Z_i\big) =P\big(N_i(C_{L_i})=0|Z_i)P\big(\mathcal{Q}_{1i}|N_i(C_{L_i})=0,Z_i\big)+P\big(N_i(C_{L_i})>0|Z_i)P\big(\mathcal{Q}_{1i}|N_i(C_{L_i})>0,Z_i\big)$.
After some simplification, $P(N_i(a-)=0|\mathcal{Q}_{1i})
    =\lambda_{01}(a_{i1})e^{\beta_1^{'}Z_i}\exp\{-\int_{0}^{a_{i1}}\lambda_{01}(u)e^{\beta_1^{'}Z_i}du\}/Dr$, where $Dr=\lambda_{01}(a_{i1})e^{\beta_1^{'}Z_i}\exp\{-\int_{0}^{a_{i1}}\lambda_{01}(u)e^{\beta_1^{'}Z_i}du\}
+\lambda_{02}(a_{i1})e^{\beta_2^{'}Z_i} 
\bigg[1-\exp\{-\int_{0}^{C_{L_i}}$ $\lambda_{01}(u)e^{\beta_1^{'}Z_i}du\}\bigg]
\exp\{-\int_{C_{L_i}}^{a_{i1}}\lambda_{02}(u)e^{\beta_2^{'}Z_i}du\}$.

Given $P\bigl(Y_i^{(1)}(a)=1|\mathcal{Q}_{1i}\bigr)$, it is straightforward to obtain 
    $P\bigl(Y_i^{(2)}(a)=1|\mathcal{Q}_{1i}\bigr)
    =1-P\bigl(Y_i^{(1)}(a)=1|\mathcal{Q}_{1i}\bigr)$.

\section{Asymptotic Proofs for Proposition 2 in Section 4}
\label{sec:Appendix_proofs} 

The following conditions are sufficient for establishing the consistency and asymptotic normality of the coefficient estimators derived in Section 4 under Model (\ref{eq:model}).
\begin{enumerate}[I.]
    \item $\{N_i(\cdot),Z_i,B_i\}$ for $i=1,\cdots,|\mathcal{O}|$ are independent and identically distributed;
    \item $\lambda_{0s}(a)>0$ for $a\in (0,A^\star)$ and $s\in\mathcal{S}$;
    \item $N(A^{\star})$ is bounded by a constant with probability 1;
    \item $S(\cdot)$ is a left-continuous step function over $(0, A^\star)$; 

    \item 
    $\widehat{P}(Y^{(c)}(a)=1,Z=z) \xrightarrow{a.s.}P(Y^{(c)}(a)=1,Z=z)$ for $a \in (0,A^\star)$ and $z\in \mathcal{Z}$, where $\widehat{P}(Y^{(c)}(a)=1,Z=z)$ is the estimated probability based on some supplementary information.
\end{enumerate}

We provide a step-by-step outline of the proofs for Proposition 2 in Section \ref{sec:censoring_P_avail_S}, as detailed below. 


\subsubsection*{B.1 An asymptotic proof for the case where $\boldsymbol{S}_i(\cdot)$ is fully known and $Z_i$ and $B_i$ are available for all $i\in\mathcal{O}$} 

Under Model (\ref{eq:model}), the partial score function of $\beta_s$ based on an ideal dataset $\mathcal{Q}=\bigcup_{i\in\mathcal{O}} \mathcal{Q}_{i}=\bigcup_{i\in\mathcal{O}}\big\{\{dN_i(a):C_{L_i}<a\leq C_{R_i}\}\bigcup_{}^{}\{Z_i\}\big\}$ is 
\begin{align*}
    U_{s}(\boldsymbol{\beta})=\sum_{i\in\mathcal{O}_1}^{}\int_{0}^{A^{\star}}Y_i^{(s)}(a)\big\{Z_i-\bar{Z}_s(\beta_s;a)\big\}  Y_i^{(c)}(a)dN_i(a),
\end{align*}
for $s\in\mathcal{S}$, where $\bar{Z}_s(\beta_s;a)=G_s^{(1)}(\beta_s;a)/G_s^{(0)}(\beta_s;a)$, where $G_s^{(q)}(\beta_s;a)=$ 
 $\sum_{i\in\mathcal{O}}^{}$ $ Y_i^{(s)}(a)Y_i^{(c)}(a)$ $Z_i^{\otimes q}\exp\{\beta_s^{'}Z_i\}$ for $q=0, 1, 2$. The solution of $U_{s}(\boldsymbol{\beta})=\mathbf{0}$ is the maximum partial likelihood estimator (MPLE) of $\beta_s$, denoted by $\widehat{\beta}_s$. In the ideal case, we consider a fully known $\boldsymbol{S}(\cdot)$. In these settings, we can derive the following condition as an additional condition used in the following proofs:
\begin{itemize}     
    \item[VI.] The matrix $K_s(\beta_{0s};a)=g_s^{(2)}(\beta_{0s};a)-g_s^{(1)}(\beta_{0s};a)^{\otimes 2}/g_s^{(0)}(\beta_{0s};a)$ with $g_s^{(q)}$ $(\beta_s;a)=E\big[ Y^{(s)}(a)Y^{(c)}(a)$ $Z^{\otimes q}\exp\{\beta_s^{'}Z\}\big]$ for $q=0, 1, 2$ is positive definite for $a\in (0,A^\star)$ and $s\in\mathcal{S}$, where $\beta_{0s}$ is the true parameter.
\end{itemize}   

\noindent \textit{Proof}.
Let $D_s(a,Z)= Y^{(s)}(a)Y^{(c)}(a)\exp\{\beta_{0s}^{'}Z\}$. The matrix $K_s(\beta_{0s};a)$ can be written as
        $K_s(\beta_{0s};a)=\big\{E[ D_s(a,Z)]E[D_s(a,Z) Z^{\otimes 2}]-E[D_s(a,Z) Z]^{\otimes 2}\big\}/$ $E[D_s(a,Z)]$.
Let $b$ be a nonzero real column vector, independent of the data. To prove that $K_s(\beta_{0s};a)$ is positive definite, we need to show that $b'K_s(\beta_{0s};a)b>0$. Since $E[D_s(a,Z)]>0$ for a well-defined matrix, it is sufficient to verify that $E[ D_s(a,Z)]$ $E[D_s(a,Z) (b'Z)^2]-E[D_s(a,Z) b'Z]^2>0$. Consider a random vector $W$ that is independent of $Z$ but follows the same distribution. Then,
\begin{align*}
    \begin{split}
        &E[ D_s(a,Z)]E[D_s(a,Z) (b'Z)^2]-E[D_s(a,Z) b'Z]^2\\
       &=E[ D_s(a,W)]E[D_s(a,Z) (b'Z)^2]-E[D_s(a,W) b'W]E[D_s(a,Z) b'Z]\\
       &=\frac{1}{2}\Big\{E[ D_s(a,W)]E[D_s(a,Z) (b'Z)^2]+E[ D_s(a,Z)]E[D_s(a,W) (b'W)^2]\\
       &\hspace{0.5cm}-2E[D_s(a,W) b'W]E[D_s(a,Z) b'Z] \Big\}\\
       &=\frac{1}{2}E\big[ D_s(a,W)D_s(a,Z)(b'Z-b'W)^2\big]\geq 0.
    \end{split}
\end{align*}
Since $Z$ cannot always be equal to $W$, we have $E[ D_s(a,Z)]E[D_s(a,Z) (b'Z)^2]$ $-E[D_s(a,Z) b'Z]^2>0$. Thus, $K_s(\beta_{0s};a)$ is a positive definite matrix. \qedsymbol{}
\bigskip

Extending \cite{AGmodel}, we can show $\widehat{\beta}_s$ has the following properties under conditions (I)-(V). 
 
 \begin{enumerate}[(i)]
    \item strong consistency: $\widehat{\beta}_s \xrightarrow{a.s.} \beta_{0s}$ as $n\rightarrow \infty$, and
     \item asymptotic normality: $\sqrt{n}(\widehat{\beta}_s-\beta_{0s})\xrightarrow{d} N \big(\mathbf{0},\Sigma_s(\beta_{0s})^{-1}\big)$ as $n\rightarrow \infty$, where $n=|\mathcal{O}|$.
 \end{enumerate}
 The consistent estimator of $\Sigma_s(\beta_{0s})$ is $-n^{-1}\partial U_{s}(\boldsymbol{\widehat{\beta}})/\partial \beta_{s}$, where $\boldsymbol{\widehat{\beta}}=\{\widehat{\beta}_{s}:s\in\mathcal{S}\}$. Similarly, we can show that the cumulative baseline estimator \citep{breslow_est1972}, 
\begin{align*}
    \widehat{\Lambda}_{0s}(a\big|\widehat{\boldsymbol{\beta}})=\sum_{i\in \mathcal{O}_1} \int_{0}^{a} \frac{Y_i^{(s)}(u)}{G_s^{(0)}(\widehat{\beta}_s;u)}Y_i^{(c)}(u)dN_i(u),
\end{align*}
 converges uniformly to the true baseline $\Lambda_{0s}(a)$ for $a\in(0,A^\star)$ and $\sqrt{n}\{\widehat{\Lambda}_{0s}(\cdot\big|\widehat{\boldsymbol{\beta}})-\Lambda_{0s}(\cdot)\}$ converges weakly to a zero-mean Gaussian process \citep{AGmodel}.

\subsubsection*{B.2 An asymptotic proof for the case where $\boldsymbol{S}_i(\cdot)$ is fully known for $i\in\mathcal{O}_1$, along with $Z_i$ and $B_i$, population census information} 
\noindent In addition to the conditions above, we can derive the following additional condition:
\begin{itemize}
     
        \item[VII.] The estimator $\tilde{\Lambda}_{0s}(\cdot)$ converges uniformly to $\Lambda_{0s}(\cdot)$ for $s\in\mathcal{S}$. As $n \rightarrow \infty$, $\sqrt{n}\big(\tilde{\Lambda}_{0s}(a)-\Lambda_{0s}(a)\big)\xrightarrow{w.}\mathcal{G}(a)$ with $a\in(0,A^\star)$ and $s\in \mathcal{S}$, where $\mathcal{G}(a)$ is a Gaussian process with mean zero and $n=|\mathcal{O}|$. 
\end{itemize}

When the covariates and censoring times are only known for subjects in $\mathcal{O}_1$, we consider the following estimating function (EF): 
\begin{align*}
\tilde{U}_{s}(\boldsymbol{\beta})=\sum_{i\in \mathcal{O}_1}^{}\int_{0}^{A^{\star}}Y_i^{(s)}(a)\big\{Z_i-\tilde{\bar{Z}}_s(\beta_s;a)\big\}  Y_i^{(c)}(a)dN_i(a),
\end{align*}
where $\tilde{\bar{Z}}_s(\beta_s;a)=\tilde{G}_s^{(1)}(\beta_s;a)/\tilde{G}_s^{(0)}(\beta_s;a)$ with 
\begin{equation*}
    \tilde{G}_s^{(q)}(\beta_s;a)=\sum_{z\in \mathcal{Z}}^{}\Bigl\{
P\bigl(Y^{(s)}(a)=1|Z=z\bigr)z^{\otimes q}\exp\{\beta_s^{'}z\}\bigl[\sum_{l}\mathcal{C}(l,z, a )\bigr]\Bigr\},
\end{equation*}
for $s\in\mathcal{S}$. In the proof, we re-write $\tilde{U}_{s}(\boldsymbol{\beta})$ as $\tilde{U}_{s}(\boldsymbol{\beta}|\boldsymbol{\lambda}_{0}(\cdot))$ to address that $\tilde{G}_s^{(q)}(\beta_s;a)$ could include $\boldsymbol{\lambda}_{0}(\cdot)=\{\lambda_{0s}(\cdot):s\in\mathcal{S}\}$  and 
 $\boldsymbol{\beta}=\{\beta_s:s\in\mathcal{S}\}$. Let $\tilde{U}(\boldsymbol{\beta}|\boldsymbol{\lambda}_{0}(\cdot))=\big(\tilde{U}_{1}(\boldsymbol{\beta}|\boldsymbol{\lambda}_{0}(\cdot))',\cdots, \tilde{U}_{S^\star}(\boldsymbol{\beta}|\boldsymbol{\lambda}_{0}(\cdot))'\big)'$, where $\mathcal{S}=(1,\cdots, S^\star)$. 
 
 Given the true baseline $\boldsymbol{\lambda}_{0}(\cdot)$, we obtain the estimator $\widetilde{\boldsymbol{\beta}}^\star=\{\tilde{\beta}_s^\star:s\in\mathcal{S}\}$ by solving the estimating equation (EQ) $\tilde{U}(\boldsymbol{\beta}|\boldsymbol{\lambda}_{0}(\cdot))=\mathbf{0}$ for $\boldsymbol{\beta}$. Under condition (V), $\tilde{G}_s^{(q)}(\beta_s;a)/n\xrightarrow{a.s.}g_s^{(q)}(\beta_{s};a)$ 
 for $s\in\mathcal{S}$. Thus, $\tilde{\bar{Z}}_s(\beta_s;a)$ has the same limit with $\bar{Z}_s(\beta_s;a)$ in Case (B.1). Under condition (I), $n^{-1}\tilde{U}(\boldsymbol{\beta}_0|\boldsymbol{\lambda}_{0}(\cdot))\xrightarrow{a.s.}\mathbf{0}$ and $n^{-1}\partial\tilde{U}(\boldsymbol{\beta}_0|\boldsymbol{\lambda}_{0}(\cdot))/\partial \boldsymbol{\beta}\xrightarrow{a.s.}\Pi^\star(\boldsymbol{\beta}_0)$ as $n\rightarrow \infty$ by the strong law of large numbers (SLLN), where $\boldsymbol{\beta}_0$ represents true parameters. 
 Under condition (VI), $-\Pi^\star(\boldsymbol{\beta}_0)$ is positive definite. 
 By the central limit theorem (CLT), we can show $n^{-1/2}\tilde{U}(\boldsymbol{\beta}_0|\boldsymbol{\lambda}_{0}(\cdot))\xrightarrow{d} N \big(\mathbf{0}, \Sigma^\star(\boldsymbol{\beta}_{0})\big)$ as $n\rightarrow \infty$. 
 By \cite{yuan1998asymptotics}, we can derive the strong consistency and asymptotic normality of $\widetilde{\boldsymbol{\beta}}^\star$.

 When $\boldsymbol{\lambda}_{0}(\cdot)$ is unknown and is estimated by $\widetilde{\boldsymbol{\lambda}}_{0}(\cdot)=\{\tilde{\lambda}_{0s}(\cdot):s\in\mathcal{S}\}$ satisfying the condition (VII), we solve $\tilde{U}(\boldsymbol{\beta}|\widetilde{\boldsymbol{\lambda}}_{0}(\cdot))=\mathbf{0}$ and obtain the estimator $\widetilde{\boldsymbol{\beta}}$. By the Taylor expansion of $\tilde{U}(\widetilde{\boldsymbol{\beta}}|\widetilde{\boldsymbol{\lambda}}_0(\cdot))$ in a neighborhood of $\boldsymbol{\beta}_0$, we have

\begin{align*}
   \frac{1}{n}\tilde{U}(\widetilde{\boldsymbol{\beta}}|\widetilde{\boldsymbol{\lambda}}_0(\cdot))&= \frac{1}{n}\tilde{U}(\boldsymbol{\beta}_0|\widetilde{\boldsymbol{\lambda}}_0(\cdot))+ \frac{1}{n}\frac{\partial \tilde{U}(\boldsymbol{\beta}_0|\widetilde{\boldsymbol{\lambda}}_0(\cdot))}{\partial \boldsymbol{\beta}}(\widetilde{\boldsymbol{\beta}}-\boldsymbol{\beta}_0)+o\big((\widetilde{\boldsymbol{\beta}}-\boldsymbol{\beta}_0)\big)\\
   &= \frac{1}{n}\tilde{U}(\boldsymbol{\beta}_0|\boldsymbol{\lambda}_0(\cdot))+ \frac{1}{n}\Big[\tilde{U}(\boldsymbol{\beta}_0|\widetilde{\boldsymbol{\lambda}}_0(\cdot))-\tilde{U}(\boldsymbol{\beta}_0|\boldsymbol{\lambda}_0(\cdot))\Big]\\
   &+ \frac{1}{n}\frac{\partial \tilde{U}(\boldsymbol{\beta}_0|\boldsymbol{\lambda}_0(\cdot))}{\partial \boldsymbol{\beta}}(\widetilde{\boldsymbol{\beta}}-\boldsymbol{\beta}_0)+ \frac{1}{n}\Bigg[\frac{\partial \tilde{U}(\boldsymbol{\beta}_0|\widetilde{\boldsymbol{\lambda}}_0(\cdot))}{\partial \boldsymbol{\beta}}\\
   &-\frac{\partial \tilde{U}(\boldsymbol{\beta}_0|\boldsymbol{\lambda}_0(\cdot))}{\partial \boldsymbol{\beta}}\Bigg](\widetilde{\boldsymbol{\beta}}-\boldsymbol{\beta}_0)+o\big((\widetilde{\boldsymbol{\beta}}-\boldsymbol{\beta}_0)\big).
\end{align*}
Following the proof of Lemma 1 in \cite*{LiDongdong2020asymptotic}, we can show that $\big|\tilde{U}(\boldsymbol{\beta}_0|\widetilde{\boldsymbol{\lambda}}_0(\cdot))-\tilde{U}(\boldsymbol{\beta}_0|\boldsymbol{\lambda}_0(\cdot))\big|/n \xrightarrow{a.s.}\mathbf{0}$, $\big|\partial \tilde{U}(\boldsymbol{\beta}_0|\widetilde{\boldsymbol{\lambda}}_0(\cdot))/\partial \boldsymbol{\beta}-\partial \tilde{U}(\boldsymbol{\beta}_0|\boldsymbol{\lambda}_0(\cdot))/$ $\partial \boldsymbol{\beta}\big|/n \xrightarrow{a.s.}\mathbf{0}$, and $n^{-1/2}\tilde{U}(\boldsymbol{\beta}_0|\widetilde{\boldsymbol{\lambda}}_0(\cdot))\xrightarrow{d} N \big(\mathbf{0}, \Sigma^{\star\star}(\boldsymbol{\beta}_{0})\big)$ as $n\rightarrow \infty$, which yields that $\widetilde{\boldsymbol{\beta}}$ has the following properties:

 
 \begin{enumerate}[(i)]
    \item strong consistency: $\widetilde{\boldsymbol{\beta}} \xrightarrow{a.s.}\boldsymbol{\beta}_0$ as $n\rightarrow \infty$, 
    and
     \item asymptotic normality: $\sqrt{n}(\widetilde{\boldsymbol{\beta}}-\boldsymbol{\beta}_0)\xrightarrow{d} N \big(\mathbf{0}, AV^\star(\boldsymbol{\beta}_0)\big)$ as $n\rightarrow \infty$.
 \end{enumerate}
 The asymptotic variance $AV^\star(\beta_{0s})$ can be estimated using a resampling method, similar to the method presented in Section 4.2.

 The cumulative baseline estimator \citep{breslow_est1972} is
\begin{align*}
    \tilde{\Lambda}_{0s}(a\big|\widetilde{\boldsymbol{\beta}})=\int_{0}^{a}\sum_{i\in \mathcal{O}_1} \frac{Y_i^{(s)}(u)}{\tilde{G}_s^{(0)}(\tilde{\beta}_s;u)}Y_i^{(c)}(u)dN_i(u).
\end{align*}
Under condition (I), 
$n^{-1}$ $\sum_{i=1}^n Y_i^{(s)}(u)Y_i^{(c)}(u)dN_i(u)\xrightarrow{a.s.}E[Y_1^{(s)}(u)Y_1^c(u)dN_1(u)]$ $=g_s^{(0)}(\beta_{0s};u)d\Lambda_{0s}(u)$ as $n\rightarrow \infty$ by SLLN. Under condition (V), $\tilde{G}_s^{(0)}(\beta_s;a)/n\xrightarrow{a.s.}g_s^{(0)}(\beta_{s};a)$ 
for $s\in\mathcal{S}$. Then, $\tilde{\Lambda}_{0s}(a\big|\widetilde{\boldsymbol{\beta}})$ uniformly converges to $\int_{0}^{a} \{g_s^{(0)}(\beta_{0s};u)/g_s^{(0)}(\tilde{\beta}_{s};$ $u)\}d\Lambda_{0s}(u)$ for $a\in[0, A^\star)$. The strong consistency of $\widetilde{\boldsymbol{\beta}}$ implies that $\tilde{\Lambda}_{0s}(\cdot\big|\widetilde{\boldsymbol{\beta}})$ converges uniformly to $\Lambda_{0s}(\cdot)$ in $a$.

The Taylor expansion of $\tilde{\Lambda}_{0s}(a\big|\widetilde{\boldsymbol{\beta}})$ about $\boldsymbol{\beta_0}$ is
\begin{align}
\begin{split}
   \tilde{\Lambda}_{0s}(a\big|\widetilde{\boldsymbol{\beta}})=\tilde{\Lambda}_{0s}(a\big|\boldsymbol{\beta}_0)+\frac{\partial  \tilde{\Lambda}_{0s}(a\big|\boldsymbol{\beta}_0) }{\partial \boldsymbol{\beta}}(\widetilde{\boldsymbol{\beta}}-\boldsymbol{\beta}_0)+o\big((\widetilde{\boldsymbol{\beta}}-\boldsymbol{\beta}_0)\big).
\end{split}
\end{align}
To prove the weak convergence of $\tilde{\Lambda}_{0s}(a\big|\widetilde{\boldsymbol{\beta}})$, we start with a simple decomposition:
\begin{align}
\begin{split}
\label{eq:expand_baseline}
    \sqrt{n}\{\tilde{\Lambda}_{0s}(a\big|\widetilde{\boldsymbol{\beta}})-\Lambda_{0s}(a)\}&=\sqrt{n}\{\tilde{\Lambda}_{0s}(a\big|\widetilde{\boldsymbol{\beta}})-\tilde{\Lambda}_{0s}(a\big|\boldsymbol{\beta_0})\}+\sqrt{n}\{\tilde{\Lambda}_{0s}(a\big|\boldsymbol{\beta_0})-\Lambda_{0s}(a)\}\\
    &=\frac{\partial  \tilde{\Lambda}_{0s}(a\big|\boldsymbol{\beta}_0) }{\partial \boldsymbol{\beta}}\sqrt{n}(\widetilde{\boldsymbol{\beta}}-\boldsymbol{\beta}_0)+\sqrt{n}\{\tilde{\Lambda}_{0s}(a\big|\boldsymbol{\beta_0})-\Lambda_{0s}(a)\}\\
    &\hspace{0.5cm}+o\big((\widetilde{\boldsymbol{\beta}}-\boldsymbol{\beta}_0)\big).
\end{split}
\end{align}
Under conditions (I) and (V), $\partial  \tilde{\Lambda}_{0s}(a\big|\boldsymbol{\beta}_0)/\partial \boldsymbol{\beta}$ converges almost surely to a function of $a$, $h(a|\boldsymbol{\beta}_0)$. 
Due to the asymptotic normality of $\widetilde{\boldsymbol{\beta}}$, we can show that the first term in Equation (\ref{eq:expand_baseline}) follows a Gaussian process with mean zero. Following \cite*{rate_LinD.Y.2000Srft}, let $dM_{i}(a)=Y_i^{(s)}(a)Y_i^{(c)}(a)[dN_i(a)-exp\{\beta_s^{'}Z_i\}d\Lambda_{0s}(a)]$. The second term is asymptotically equal to

\begin{align*}
 n^{-1/2}\sum_{i\in \mathcal{O}_1}\int_{0}^{a} \frac{dM_i(u)}{g_s^{(0)}(\beta_{0s};u)},
\end{align*}
which also follows a Gaussian process with mean zero. Therefore, $\sqrt{n}\{\tilde{\Lambda}_{0s}(a\big|\widetilde{\boldsymbol{\beta}})-\Lambda_{0s}(a)\}$ converges weakly to a Gaussian process with mean zero and condition (VII) is proved. 

 \subsubsection*{B.3 An asymptotic proof for the case where $\boldsymbol{S}_i(\cdot)$ is partially known for $i\in\mathcal{O}_1$, along with $Z_i$ and $B_i$, population census information} 
When the stratification variable $\boldsymbol{S}(\cdot)$ is partially known, we consider the EF:
\begin{align*}
\begin{split}
   \tilde{U}^\star_{s}(\boldsymbol{\beta})=\sum_{i\in \mathcal{O}_1}^{}\int_{0}^{A^{\star}}P\bigl(Y_i^{(s)}(a)=1|\mathcal{Q}_{1i}\bigr)\big\{Z_i-\tilde{\bar{Z}}_s(\beta_s;a)\big\}  Y_i^{(c)}(a)dN_i(a),
    \label{eq:score_fun-p_s_appx}
    \end{split}
\end{align*}
for $s\in\mathcal{S}$, where $\mathcal{Q}_1=\bigcup_{i\in\mathcal{O}_1} \mathcal{Q}_{1i}=\bigcup_{i\in\mathcal{O}_1}\big\{\{dN_i(a):C_{L_i}<a\leq C_{R_i}\}\bigcup_{}^{}\{Z_i\}\big\}$. We re-write $\tilde{U}^\star_{s}(\boldsymbol{\beta})$ as $\tilde{U}^\star_{s}(\boldsymbol{\beta}|\boldsymbol{\lambda}_0(\cdot))$. When the true $\boldsymbol{\lambda}_0(\cdot)$ is known, we obtain the estimator $\widetilde{\widetilde{\boldsymbol{\beta}}}^\star$ by solving $\tilde{U}^\star(\boldsymbol{\beta}|\boldsymbol{\lambda}_0(\cdot))=$ $\big(\tilde{U}^\star_1(\boldsymbol{\beta}|\boldsymbol{\lambda}_0(\cdot))',\cdots,\tilde{U}^\star_{S^\star}(\boldsymbol{\beta}|\boldsymbol{\lambda}_0(\cdot))'\big)'=\mathbf{0}$. Under conditions (I) and (V), we can show that $\tilde{U}^\star(\boldsymbol{\beta}|\boldsymbol{\lambda}_0(\cdot))/n \xrightarrow{a.s.}\mathbf{0}$ as $n\rightarrow \infty$ 
 by SLLN  and $\tilde{U}^\star(\boldsymbol{\beta}|\boldsymbol{\lambda}_0(\cdot))/\sqrt{n}$ converges to a multivariate normal distribution with a mean of 0 by CLT. Under conditions (I) and (VI), $n^{-1}\partial \tilde{U}^\star(\boldsymbol{\beta}|\boldsymbol{\lambda}_0(\cdot))/\partial \boldsymbol{\beta}$ converges to a positive definite matrix by the SLLN. 

When the true $\boldsymbol{\lambda}_0(\cdot)$ is unknown and is estimated by $\tilde{\tilde{\boldsymbol{\lambda}}}_0(\cdot)$ satisfying a condition similar to condition (VII), we obtain the estimator $\widetilde{\widetilde{\boldsymbol{\beta}}}$ by solving $\tilde{U}^\star(\boldsymbol{\beta}|\tilde{\tilde{\boldsymbol{\lambda}}}_0(\cdot))=\mathbf{0}$. Like Case (B.2), we consider the Taylor expansion of $\tilde{U}^\star(\widetilde{\widetilde{\boldsymbol{\beta}}}|\tilde{\tilde{\boldsymbol{\lambda}}}_0(\cdot))$ around the true coefficients and follow the work of \cite*{LiDongdong2020asymptotic} to complete the proof for Proposition 2.

\newpage

\section*{Competing interests}
No competing interest is declared.

\section*{Acknowledgements}

\par
We thank Professor Boxin Tang for his valuable assistance with the asymptotic proofs. \\
\noindent Disclaimer:
This study is based in part on data provided by Alberta Health. The interpretation and conclusions are contained herein are those of the researchers and do not necessarily represent the views of the Government of Alberta. Neither the government nor Alberta Health expresses any opinion in relation to this study.

\newpage
\bibliography{myref}

\label{lastpage}

\end{document}